\documentclass[10pt,conference]{IEEEtran}

\usepackage[switch]{lineno}
\usepackage{amsmath,amsfonts}
\usepackage[frozencache,cachedir=.]{minted}
\usepackage[linesnumbered,ruled,lined,noend]{algorithm2e}
\usepackage{algorithmic}
\usepackage{graphicx}
\usepackage{textcomp}
\usepackage[dvipsnames]{xcolor}
\usepackage[hidelinks]{hyperref}
\usepackage{import} 
\usepackage[nameinlink]{cleveref}
\usepackage{ifthen} 
\usepackage{xspace}
\usepackage{framed}
\usepackage{subcaption} 
\usepackage{listings}
\usepackage{soul} 
\usepackage{wasysym}
\usepackage{booktabs}
\usepackage{tabularx}
\usepackage{multirow}
\usepackage{pifont}
\usepackage{stmaryrd} 
\usepackage{tcolorbox}
\tcbuselibrary{skins,minted}
\usepackage{cite}
\usepackage{amssymb}
\usepackage{enumitem}
\usepackage{tikz}
\usetikzlibrary{calc,tikzmark}
\usepackage{wrapfig}

\makeatletter
\providecommand*{\Dashv}{%
  \mathrel{%
    \mathpalette\@Dashv\vDash
  }%
}
\newcommand*{\@Dashv}[2]{%
  \reflectbox{$\m@th#1#2$}%
}
\newcommand{\linebreakand}{%
  \end{@IEEEauthorhalign}
  \hfill\mbox{}\par
  \mbox{}\hfill\begin{@IEEEauthorhalign}
}
\makeatother

\newcolumntype{Y}{>{\centering\arraybackslash}X} %

\definecolor{myred}{HTML}{aa0000}
\definecolor{mygreen}{HTML}{008000}
\lstdefinelanguage{diff}{
  morecomment=[f][\color{mygreen}]{+\ },
  morecomment=[f][\color{myred}]{-\ },
}
\definecolor{gray}{RGB}{215,215,215}
\sethlcolor{gray}

\newcommand{\hide}[1]{}

\newcommand{\tname}[1]{\textsc{#1}\xspace}
\newcommand{\tool}{{\sc Silc}}
\newcommand{\pulse}{\tname{Pulse}}

\definecolor{light-gray}{gray}{0.90}
\definecolor{dark-gray}{gray}{0.40}
\definecolor{light-red}{HTML}{FFCCCB}
\definecolor{light-yellow}{HTML}{FDEFB2}
\definecolor{deepblue}{rgb}{0,0,0.5}
\definecolor{deepred}{rgb}{0.6,0,0}
\definecolor{deepgreen}{rgb}{0,0.5,0}

\definecolor[named]{ACMBlue}{cmyk}{1,0.1,0,0.1}
\definecolor[named]{ACMYellow}{cmyk}{0,0.16,1,0}
\definecolor[named]{ACMOrange}{cmyk}{0,0.42,1,0.01}
\definecolor[named]{ACMRed}{cmyk}{0,0.90,0.86,0}
\definecolor[named]{ACMLightBlue}{cmyk}{0.49,0.01,0,0}
\definecolor[named]{ACMGreen}{cmyk}{0.20,0,1,0.19}
\definecolor[named]{ACMPurple}{cmyk}{0.55,1,0,0.15}
\definecolor[named]{ACMDarkBlue}{cmyk}{1,0.58,0,0.21}

\hypersetup{colorlinks,
  linkcolor=ACMDarkBlue,
  citecolor=ACMPurple,
  urlcolor=ACMDarkBlue,
  filecolor=ACMDarkBlue}

\newcommand{\errstep}[1]{\scriptsize\colorbox{yellow}{\bf(#1)}}

\newcommand{\constext}[1]{\mathsf{#1}}
\newcommand{\mallocPCword}{\constext{malloc}}
\newcommand{\freePCword}{\constext{free}}

\newcommand{\True}{\mathsf{True}}

\newcommand{\mcode}[1]{{\ensuremath{\tt #1}}}
\newcommand{\fnc}{\tt {fnc}}
\newcommand{\algo}[1]{\textsc{#1}}

\newtcblisting{boxabridged}{
    rounded corners, 
    boxrule = 0pt,
    toprule = 0pt, 
    enhanced,
    drop fuzzy shadow = black!35,
    fontupper=\sffamily,
    listing only,
    minted language=C,
    minted options={
    xleftmargin=-1em,
    fontsize=\scriptsize,}
}

\newtcblisting{boxprompt}{
    sharpish corners, 
    boxrule = 0pt,
    toprule = 0pt, 
    enhanced,
    drop fuzzy shadow = black!35,
    fontupper=\sffamily,
    listing only,
    minted language=C,
    minted options={
    xleftmargin=-1em,
    fontsize=\scriptsize,
    escapeinside=||,}
}

\newcommand{\tclient}{{\tt{Client}}}
\newcommand{\tvendor}{{\tt{Vendor}}}
\newcommand{\tblame}{{\tt{Blame}}}
\newcommand{\tworld}{{\tt{World}}}
\newcommand{\cworld}{\color{Fuchsia}{\tworld}}
\newcommand{\cblame}{\color{orange}{\tblame}}

\newcommand{\pworld}[4]{\tworld(#1,#2,#3)}
\newcommand{\pblame}[5]{\tblame(#1,#2,#3,#4)}

\newcommand{\world}[4]{{\color{Fuchsia}{\pworld{#1}{#2}{#3}{#4}}}}
\newcommand{\blame}[5]{{\color{orange}{\pblame{#1}{#2}{#3}{#4}{#5}}}}

\newcommand{\errprot}[2]{#1 \implies #2}
\newcommand{\bugdef}[2]{#1 \implies #2}
\newcommand{\bug}[2]{#1(#2)}
\newcommand{\bugcond}{\beta}
\newcommand{\sanitise}[2]{#1 \implies #2}

\newcommand{\pts}{\mapsto}
\newcommand{\pointsto}[2]{#1 \pts #2}
\newcommand{\dealloc}[1]{#1 \not\pts}

\newcommand{\islentails}[2]{{#1}\Dashv{#2}}

\newcommand{\emp}{\mathsf{emp}}
\newcommand{\sep}{\ast}
\newcommand{\nil}{\mathsf{nil}}
\newcommand{\isltriple}[3]{[{{\color{ACMBlue} \tt #1}}]~{\tt #2}~[{\color{ACMBlue}  #3}]}
\newcommand{\okisltriple}[3]{[{{\color{ACMBlue} \tt #1}}]~{\tt #2}~[{\color{deepgreen}\ok:  #3}]}
\newcommand{\errisltriple}[3]{[{{\color{ACMBlue} \tt #1}}]~{\tt #2}~[{\color{ACMRed} \err: #3}]}
\newcommand{\errblameisltriple}[4]{[{{\color{ACMBlue} \tt #1}}]~{\tt #2}~[{\color{ACMRed} \err^{#4}: #3}]}

\newcommand{\greenstate}[1]{{\color{deepgreen} #1}}
\newcommand{\redstate}[1]{{\color{ACMRed} #1}}
\newcommand{\bluestate}[1]{{\color{ACMBlue} #1}}

\newcommand{\errlbl}{\epsilon}
\newcommand{\ok}{\tt{ok}}
\newcommand{\err}{\tt{err}}
\newcommand{\cok}{{\color{deepgreen}\ok}}
\newcommand{\cerr}{{\color{ACMRed}\err}}

\newcommand{\oldnew}[2]{\ifthenelse{#1>0}{{\textcolor{red}{#1}}}{#1}+\ifthenelse{#2>0}{{\textcolor{red}{#2}}}{#2}}
\newcommand{\manifest}[1]{\ifthenelse{#1>0}{{\textcolor{red}{#1}}}{#1}}
\newcommand{\saneval}[1]{\ifthenelse{\equal{#1}{100}}{{\textcolor{deepgreen}{#1\%}}}{\ifthenelse{\equal{#1}{0}}{{\textcolor{red}{#1\%}}}{\textcolor{ACMGreen}{#1\%}}}}
\newcommand{\gptthree}{GPT-3.5}
\newcommand{\gptfour}{GPT-4o}
\newcommand{\gpt}{GPT}
\newcommand{\na}{\textcolor{dark-gray}{N.A.}}
\colorlet{MLsafecolour}{Salmon!50}
\colorlet{MLunsafecolour}{Salmon!100}
\colorlet{NPDsafecolour}{cyan!50}
\colorlet{NPDunsafecolour}{cyan!100}
\colorlet{UAFsafecolour}{Goldenrod!50}
\colorlet{UAFunsafecolour}{Goldenrod!110}

\newcommand{\slice}[5]{
  \pgfmathparse{0.5*#1+0.5*#2}
  \let\midangle\pgfmathresult

  \draw[thick,fill=#5,line width=2pt,draw=white] (0,0) -- (#1:1) arc (#1:#2:1) -- cycle;

  \node[label=\midangle:#4] at (\midangle:0.8) {};

  \pgfmathparse{min((#2-#1-10)/110*(-0.3),0)}
  \let\temp\pgfmathresult
  \pgfmathparse{max(\temp,-0.5) + 0.8}
  \let\innerpos\pgfmathresult
  \node at (\midangle:\innerpos) {#3};
}

\newcommand{\slicenolbl}[5]{
  \pgfmathparse{0.5*#1+0.5*#2}
  \let\midangle\pgfmathresult

  \draw[thick,fill=#5,line width=0pt,draw=#5] (0,0) -- (#1:1) arc (#1:#2:1) -- cycle;
}

\newcommand{\slicefull}[5]{
  \pgfmathparse{0.5*#1+0.5*#2}
  \let\midangle\pgfmathresult

  \pgfmathparse{0.99*#1+0.01*#2}
  \let\rightangle\pgfmathresult

   \draw[thick,fill=#5,line width=0pt,draw=#5] (0,0) -- (#1:1) arc (#1:#2:1) -- cycle;

  \node[label=\midangle:#4] at (\rightangle:0.8) {};
}

\newcommand{\sliceline}[2]{
  \pgfmathparse{0.5*#1+0.5*#2}
  \let\midangle\pgfmathresult

  \draw[thick,fill=none,line width=2pt,draw=white] (0,0) -- (#1:1) arc (#1:#2:1) -- cycle;
}

\newcommand{\innerlabel}[3]{
  \pgfmathparse{0.5*#1+0.5*#2}
  \let\midangle\pgfmathresult

  \pgfmathparse{min((#2-#1-10)/110*(-0.3),0)}
  \let\temp\pgfmathresult
  \pgfmathparse{max(\temp,-0.5) + 0.8}
  \let\innerpos\pgfmathresult
  \node at (\midangle:\innerpos) {#3};
}

\begin{document}

\title{Whose fault is it anyway? \\ \tool: \textbf{S}afe \textbf{I}ntegration of \textbf{L}LM-Generated \textbf{C}ode}

\author{
    \IEEEauthorblockN{Peisen Lin}
    \IEEEauthorblockA{
    National University of Singapore\\
    Singapore\\
    peisen-l@comp.nus.edu.sg}
    \and
    \IEEEauthorblockN{Yuntong Zhang}
    \IEEEauthorblockA{
    National University of Singapore\\
    Singapore\\
    yuntong@comp.nus.edu.sg}
    \linebreakand
    \IEEEauthorblockN{Andreea Costea}
    \IEEEauthorblockA{
    National University of Singapore\\
    Singapore\\
    andreeac@comp.nus.edu.sg}
    \and
    \IEEEauthorblockN{Abhik Roychoudhury}
    \IEEEauthorblockA{
    National University of Singapore\\
    Singapore\\
    abhik@comp.nus.edu.sg}
}
\maketitle
\begin{abstract}
In modern software development, multiple software components, often sourced from different contributors, including AI assistants, are combined to create a cohesive system.
Although these components might each be individually safe, their composition might not be so. 
At the core of this issue is often a misalignment between the requirements and assumptions made by each component. 
Once discovered it is important to determine which component is accountable for addressing the misalignment issue and to prevent its occurrence in the future.

In this work we propose \tool, a framework for localising fault, i.e. blame, and  for assigning sanitization obligations to prevent memory issues resulting from the composition of multiple software components.
In particular, we show the role Incorrectness Logic could have in automatically extracting implicit non-functional assumptions in auto-generated code and render them explicit in order to detect misalignment with the requirements in existing code.
In other words, we are looking at the problem of code comprehension from a perspective focused on safety properties rather than the traditional approach centered on functionality. 
To do that, we enhance Incorrectness Separation Logic with capabilities for 
fault tracking and sanitization insertion. 
We show the benefits of this framework by running experiments on millions of lines of code from open source projects where parts of existing functionality are regenerated by AI assistants. We empirically show that AI assistants produce unsafe code and demonstrate the utility of our framework in proposing appropriate blame and sanitization obligations.

\end{abstract}


\maketitle

\section{Introduction}
\label{sec:introduction}


Generative AI is disrupting the way we design, build and maintain software systems. 
Assistants based on Large Language Model (LLM), such as Copilot \cite{githubCopilot}, are becoming integral part of the software development workflow. 
They are used for code debugging and bug finding \cite{Haonan2024,Sun2024,meng2024large},
for automated program repair \cite{fan2023automated,Chunqiu2023,Joshi2023}, and code translation \cite{szafraniec2023code,Baptiste2020,eniser2024translating}, among other LLM applications that have garnered significant research interest \cite{puri2021codenet}.
The appeal of LLM for code generation is by now evident: even without much model fine-tuning, a developer only provides a brief description of the task in natural languages and the LLM completes the task immediately. 
The impact of this technology is remarkable \cite{CopilotImpact2024}.

However, there is a flip-side to this technological goodness. Developers may need to debug and integrate code into existing software that they do not fully understand\cite{AIProgrammingAssistants2024,ICSEcopilot2022,GILT2024}.   
%
When bugs appear after the integration of auto-generated code into existing projects, it is often difficult to say what their source is \cite{expectations2022}. 
It is often unclear whether they appear because the auto-generated code is incorrect, unsafe or insecure, or because there is a misalignment between the requirements of the existing code and the assumptions implicitly contained within the auto-generated code.
While there are efforts for building technologies to check for the functional correctness of auto-generated code \cite{Jiawei2024,misu2024towards},
there is no research indicating how to assign repair obligations, i.e. \emph{blame} and \emph{sanitisation}, 
for non-functional bugs arising after the integration of the auto-generated code into existing code. 
%
%
What we aim to investigate in this work is how to enhance the code with abstractions that carry blame and sanitization obligations. This would ensure that the software remains safe even as it evolves with the inclusion of auto-generated code.
That is, we are looking at how to reconcile the misalignment between the non-functional requirements of the existing code and the implicit assumptions built within the auto-generated code. 
This becomes a crucial mission since the development of software components is increasingly being outsourced to AI assistants which, despite their impressive effectiveness, still produce vulnerable code even when properly finetuned \cite{CyberSecEval2,surveysecurity2023}.

To address this problem, we propose \tool, a logical framework which assigns repair responsibilities and derives sanitizations to mitigate the effects of identified misalignments; sanitization
%
avoids erroneous 
executions without modifying the auto-generated code even after integration into existing codebase.
\tool\, avoids memory safety issues in C code bases, but it could potentially be extended to other kinds of non-functional requirements too. 
We base our work on the advancements of Incorrectness Separation Logic (ISL) \cite{Raad2020}.
%
ISL has been shown effective for modeling the conditions which indicate the presence of Null Pointer Dereferences, memory leaks, use-after-free (double-free is a special case of use-after-free) bugs \cite{Le2022}, and so is well suited for capturing the requirements and assumptions of the C code components \tool\, targets.
We further extend ISL with blame tracking capabilities, dubbed blame-carrying ISL.
On this logical basis, \tool\, extracts the implicit memory safety assumptions embedded in the auto-generated code and makes them explicit as ISL assertions for subsequent use.
\tool\ tracks blame and sanitization obligations when checking the requirements of existing code against the assumptions of auto-generated code. 
This ensures the detection of unsafe conditions resulting from integration,  accurately identifying the responsible component for misalignment, before applying the corresponding sanitization.
%

Our key contributions are as follows:
\begin{itemize}
    \item a blame-carrying logic that serves as a program analysis foundation to accurately identify program components responsible for introducing unsafety issues;
    \item a generic framework for assigning blame and sanitization obligations for bugs resulted from the integration of auto-generated code into existing codebases;
    \item \tool, an instantiation of this framework with support for three kinds of memory safety issues: Null Pointer Dereferences, memory leaks, use-after-free;
    \item a validation of the effectiveness of \tool\, on 19 real-world open source projects, totalling 102 distinct subjects, and 1020 auto-generated functions.
    \item two C datasets used for evaluating the effectiveness of LLMs in deriving safe code in real-world projects. 
\end{itemize}

\section{Motivating Example}
\label{sec:motivation}
\begin{figure}[t]

{\footnotesize
\begin{subfigure}{0.49\textwidth}
\colorlet{FancyVerbHighlightColor}{light-gray}
\begin{minted}[fontsize=\footnotesize,autogobble,escapeinside=||,numbers=left,breaklines=true,highlightlines={4,12,18},xleftmargin=4.0ex]{C}

void validate_client(t_client **client) {
 assert(client != NULL);
 if(/* client->cid in blocked-list */) {  |\label{ext:validate}|
   _client_list_free_node(*client);       |\label{culpritcallmemleak}|
   *client = NULL;
} } 

void client_list_add_client(t_client *client) {
 assert(client != NULL);
 t_client *new_client = (t_client *)malloc(...);
 listen_client(new_client);   |\label{ext:listen}|
 validate_client(&new_client); |\xglobal\colorlet{FancyVerbHighlightColor}{light-yellow}||\label{ext:vaidateout}|
 if(new_client){     |\label{ext:new_client}|
     while(client->next){
      client = client->next;
     }
    client->next = new_client;  
} }   |\xglobal\colorlet{FancyVerbHighlightColor}{light-red}| |\label{ext:errreport}|
\end{minted}
\caption{Programming Context (simplified for brevity)}
\vspace{-1em}
\label{fig:memleak-client}
  \end{subfigure}
}
\\~\\
{\footnotesize
\begin{subfigure}{0.49\textwidth}
\begin{boxabridged}
  Write a function which frees the memory used by the 
  client in the proper order. If the client has a CID, 
  write the client info message to remove the CID file.  
\end{boxabridged}

\caption{Prompt to implement \mintinline[fontsize=\footnotesize,autogobble,escapeinside=||,linenos=false]{C}{ _client_list_free_node} } \label{fig:memleak-prompt}
  \end{subfigure}
}
{\footnotesize
\begin{subfigure}{0.49\textwidth}
\vspace*{1em}
\renewcommand\theFancyVerbLine{%
\ifnum\value{FancyVerbLine}=6
    \tiny\setcounter{FancyVerbLine}{15}\ldots
\else
        {\tiny\arabic{FancyVerbLine}}%
\fi
}
\begin{minted}[fontsize=\footnotesize,autogobble,escapeinside=||,numbers=left,breaklines=true,xleftmargin=4.0ex,highlightlines={23}]{C}
void _client_list_free_node(t_client *client) {
 // Check if the client has a CID
 if (client->cid && strlen(client->cid) > 0) {
  char *msgbuf  = safe_calloc(SMALL_BUF);
  char *cidinfo = safe_calloc(SMALL_BUF);
   
  // Free the allocated buffers 
  free(msgbuf);  |\label{ext:local1}|
  free(cidinfo); |\label{ext:local2}|
 }
 free(client->ip); free(client->mac);  |\label{ext:external1}|
 free(client->hid); free(client->cid);
 free(client->token); 
 free(client);                         |\label{ext:external2}||\xglobal\colorlet{FancyVerbHighlightColor}{light-yellow}|
}
\end{minted}
\caption{ Auto-generated code (misses to free \mintinline[fontsize=\footnotesize]{C}{client->cpi_query})}\label{fig:memleak}
\vspace{-1em}
\end{subfigure}
}
\caption{AI-assisted programming  may lead to a memory leak.} \label{fig:exmemleak}
\vspace{-1em}
\end{figure}

Consider a scenario where a developer
is working on the implementation of the code in 
\autoref{fig:memleak-client} which is part of a larger project for a border control gateway between a public local area network and the Internet \cite{openNDS}. 
The code listens for a new client (line \ref{ext:listen}) and adds it to a list of existing clients if valid (line \ref{ext:new_client}).
If the validation fails (line \ref{ext:validate}), the client is deleted by freeing the memory used to store its details (line \ref{culpritcallmemleak})--a functionality which is yet to be added. 
The developer asks the AI assistant for help in implementing the desired functionality using a prompt similar to the one in \autoref{fig:memleak-prompt}.
In response, the AI assistant generates the code in \autoref{fig:memleak}.
The code essentially does what the developer requested and it does so with quasi-appropriate safety measures: it frees both the locally allocated memory (lines \ref{ext:local1}-\ref{ext:local2}) and that occupied by 
\mintinline[fontsize=\footnotesize]{C}{client} (lines \ref{ext:external1}-\ref{ext:external2}), freeing the structure's members before the structure itself. This is actually safer code than a developer written version of the same function which only freed the structure but not its members, thus leading to a reported memory leak in the past \cite{memleakopenNDS}.
However, what the reader might not realize unless familiar with this project is that \mintinline[fontsize=\footnotesize]{C}{t_client} has an additional field, \mintinline[fontsize=\footnotesize]{C}{cpi_query}, which the auto-generated code misses to call \mintinline[fontsize=\footnotesize]{C}{free} on. Although ``safer" than what a human developer had previously written, this function still leaks memory.

Existing bug detection technologies for memory safety issues, including static analyses \cite{Le2022} and fuzzing methods \cite{MemLock}, could uncover such issues although their root cause is not always made obvious. For example, \pulse, the static analysis we employed, was able to detect this bug, but, as highlighted in red,  reported the memory leak at line \ref{ext:errreport} (\autoref{fig:memleak-client}) instead of line \ref{ext:external2} (\autoref{fig:memleak}) or at the very least
line \ref{ext:vaidateout} (\autoref{fig:memleak-client}), as highlighted in yellow.
The root cause of the memory leak and  the steps to mitigate it remain unclear by just investigating the analysis report. 
To alleviate this issue we introduce \tool\ -- a system that traces the origin of bugs across various functions and files, 
and suggests sanitization solutions for safe code integration until a repair oracle addresses the core issues.

For the example in \autoref{fig:exmemleak}, \tool\, is designed such that it:
\begin{itemize}
    \item traces the responsibility for the memory safety issue back to \mintinline[fontsize=\footnotesize]{C}{_client_list_free_node}.
    \tool\, achieves this by keeping track of the last function that modified the data structure causing the memory leak, and of the path conditions leading to it. The mechanics of this tracking system are embedded into Incorrectness Separation Logic (ISL).  In particular, we devise a \emph{blame} predicate which is being manipulated by the ISL-based abstract state over which the static analysis \pulse operates. 
    \item creates a wrapper around the call to the culprit method and frees the leaked memory; that is, line \ref{culpritcallmemleak} is being replaced with a call to a sanitiser: 
    {
    \renewcommand\theFancyVerbLine{%
        \ifnum\value{FancyVerbLine}=1
            \tiny\setcounter{FancyVerbLine}{4}4
        \else
            \ifnum\value{FancyVerbLine}=5
            \tiny\setcounter{FancyVerbLine}{4}4
        \else
            {\tiny\arabic{FancyVerbLine}}%
        \fi
        \fi
    }
    \begin{minted}[fontsize=\footnotesize,autogobble,escapeinside=||,numbers=left,breaklines=true,xleftmargin=7.0ex]{C}
      --- _client_list_free_node(*client);
      +++ sanitise(*client);
    \end{minted}
    where \mintinline[fontsize=\footnotesize]{C}{sanitise} is generated by \tool~as:
    \begin{minted}[fontsize=\footnotesize,autogobble,escapeinside=||,numbers=none,breaklines=true,xleftmargin=4.0ex]{C}
    void sanitise(t_client *client){
      void *tmp = client->cpi_query;
      _client_list_free_node(client);
      free(tmp);
    }
    \end{minted}
    }
\end{itemize}

Another possible sanitisation solution is to simply terminate the execution of the program, or do nothing other than showing a warning to the developer. The choice of sanitisation is left to the developer, and we offer the means for the developer to define project- and bug-specific sanitisation templates. 

\begin{figure}[t]
{\footnotesize
\begin{subfigure}{0.49\textwidth}
\renewcommand\theFancyVerbLine{%
\ifnum\value{FancyVerbLine}=2
    \tiny\setcounter{FancyVerbLine}{61}\ldots
\else
\ifnum\value{FancyVerbLine}=74
     \tiny\setcounter{FancyVerbLine}{93}\ldots
\else
\ifnum\value{FancyVerbLine}=96
     \tiny\setcounter{FancyVerbLine}{147}\ldots
\else
\ifnum\value{FancyVerbLine}=151
     \tiny\setcounter{FancyVerbLine}{165}\ldots
\else
\ifnum\value{FancyVerbLine}=168
     \tiny\setcounter{FancyVerbLine}{174}\ldots
\else
\ifnum\value{FancyVerbLine}=179
     \tiny\setcounter{FancyVerbLine}{186}\ldots
\else
    {\tiny\arabic{FancyVerbLine}}%
\fi
\fi
\fi
\fi
\fi
\fi
}
\begin{minted}[fontsize=\footnotesize,autogobble,escapeinside=||,numbers=left,breaklines=true,highlightlines={62,176},xleftmargin=4.0ex]{C}
GF_Err swf_svg_show_frame(SWFReader *read){

 swf_svg_add_iso_sample(read->user,       |\errstep{3}| |\errstep{8}||\label{npdcaller}| read->svg_data, read->svg_data_size, ...); |\xglobal\colorlet{FancyVerbHighlightColor}{light-gray}|
 gf_free(read->svg_data);
 read->svg_data = NULL;                   |\errstep{5}|  
 read->svg_data_size = 0;
 read->empty_frame = GF_TRUE;
 return GF_OK;
}

GF_Err swf_process_tag(SWFReader *read){
  switch (read->tag) {
    case SWF_END:
      return GF_OK;

    case SWF_SHOWFRAME:
      return swf_show_frame(read);        |\errstep{2}| |\errstep{7}| 
  
} }

GF_Err gf_text_process_swf(GF_TXTIn *ctx,...){

 ctx->swf_parse = gf_swf_reader_new(NULL, ctx->file_name);
 assert(ctx->swf_parse != NULL)            

 while (e == GF_OK) { |\label{npd:while}|
   e = swf_process_tag(ctx->swf_parse);   |\errstep{1}| |\errstep{6}||\xglobal\colorlet{FancyVerbHighlightColor}{light-red}||\label{npdreport}|
   if (ctx->is_suspended) return GF_OK;
  }

}
\end{minted}
\vspace{-1.5em}
\caption{Calling context (simplified for brevity)}
\label{fig:npd-client}
  \end{subfigure}
}~\\
{\footnotesize
\begin{subfigure}{0.49\textwidth}
\begin{boxabridged}
  Write a function to add an ISO sample to the SVG 
  filter. [...] It then creates a new filter packet, 
  copies the sample data, [...]
\end{boxabridged}
\caption{Prompt to implement \mintinline[fontsize=\footnotesize]{C}{swf_svg_add_iso_sample} }
\label{fig:npd-prompt}
  \end{subfigure}
}
{\footnotesize
\begin{subfigure}{0.49\textwidth}
\vspace*{1em}
\renewcommand\theFancyVerbLine{%
\ifnum\value{FancyVerbLine}=2
    \tiny\setcounter{FancyVerbLine}{15}\ldots       
\else 
    \ifnum\value{FancyVerbLine}=21
       \tiny\setcounter{FancyVerbLine}{29}\ldots
    \else
    {\tiny\arabic{FancyVerbLine}}%
\fi
\fi
}
\begin{minted}[fontsize=\footnotesize,autogobble,escapeinside=||,numbers=left,breaklines=true,highlightlines={20},xleftmargin=4.0ex]{C}
GF_Err swf_svg_add_iso_sample(void *user, const u8 *data, ...) {

  GF_FilterPacket *pck;
  u8 *pck_data;
  pck = gf_filter_pck_alloc(ctx->opid, length, &pck_data);
  if (!pck) return GF_OUT_OF_MEM;
  memcpy(pck_data, data, length);         |\errstep{4}| |\errstep{9}||\xglobal\colorlet{FancyVerbHighlightColor}{light-yellow}||\label{npdreal}|

  return GF_OK;
}
\end{minted}
\caption{Auto-generated code (latent NPD at line \ref{npdreal})}
\vspace{-1em}
\label{fig:npd-auto}
  \end{subfigure}
}
\caption{AI-assisted programming may lead to a NPD.}\label{fig:npd}
\vspace{-1em}
\end{figure}

The code in \autoref{fig:npd} exemplifies a possible Null Pointer Dereference (NPD). Without diving into the functionality details of this code, we learn from 
the annotated key operations that, after a long call chain,
\mintinline[fontsize=\footnotesize]{C}{svg_data}, a member of the data structure pointed to by \mintinline[fontsize=\footnotesize]{C}{ctx->swf_parse}, \mintinline[fontsize=\footnotesize,escapeinside=||]{C}{|{\errstep{1}}|},
is passed down as an argument to the auto-generated function,
\mintinline[fontsize=\footnotesize,escapeinside=||]{C}{|{\errstep{3}}|},
which dereferences it through an invocation of \mintinline[fontsize=\footnotesize]{C}{memcpy}, 
\mintinline[fontsize=\footnotesize,escapeinside=||]{C}{|{\errstep{4}}|}.
Upon return from the call to the auto-generated function, this pointer is set to NULL, 
\mintinline[fontsize=\footnotesize,escapeinside=||]{C}{|{\errstep{5}}|}.
This subsequently results in an NPD during the second iteration of the call chain,
\mintinline[fontsize=\footnotesize,escapeinside=||]{C}{|{\errstep{6}}|},
specifically during the next invocation of \mintinline[fontsize=\footnotesize]{C}{memcpy},\mintinline[fontsize=\footnotesize,escapeinside=||]{C}{|{\errstep{9}}|}.

\pulse indeed reports an NPD at line \ref{npdreport}. However, identifying the exact problematic operation and mitigating this bug remains unclear without a tedious manual inspection like the one above. 
Often, data structures and call chains are even more complex, like those in the code we extracted this simplified example from \cite{gpac}, leaving such code vulnerable to attacks \cite{npdGPAC}. For this example, \tool:
\begin{itemize}
    \item tracks the NPD blame from line \ref{npdreport} (\autoref{fig:npd-client}) down to line \ref{npdreal} (\autoref{fig:npd-auto}). The insight in this case is that \pulse generates a method summary for \mintinline[fontsize=\footnotesize,escapeinside=||]{C}{gf_text_process_swf} abstracting the behaviour of the method. This summary captures a latent bug behavior, i.e. a possible NPD whose manifestation is dependent on the calling context:
    {\small
    \hspace*{2.5cm} $\mathsf{data} =  \mcode{NULL} \implies  \cerr  $\\
    \hspace*{2.5cm} $\mathsf{data} \neq \mcode{NULL} \implies \cok $
    }\\
    which reads ``if this method is called from a context where NULL flows into \mintinline[fontsize=\footnotesize,escapeinside=||]{C}{data}, then the program has a safety error. Else the program is safe."
    Such latent bugs are not reported to the developer, since they are often a source of false alarms.
    Our blame predicate tracks this information when composing method summaries, thus detecting that it is the latent bug of the \mintinline[fontsize=\footnotesize,escapeinside=||]{C}{swf_svg_add_iso_sample} function that becomes a manifesting bug at line \ref{npdreport}.
    \item generates a sanitiser for the call to the auto-generated function in order to avoid the buggy program path, i.e. that path which turns a latent bug into a manifesting one:
    {
    \renewcommand\theFancyVerbLine{%
        \ifnum\value{FancyVerbLine}=1
            \tiny\setcounter{FancyVerbLine}{62}62
        \else
            \ifnum\value{FancyVerbLine}=63
            \tiny\setcounter{FancyVerbLine}{62}62
        \else
            {\tiny\arabic{FancyVerbLine}}%
        \fi
        \fi
    }
    \begin{minted}[fontsize=\footnotesize,autogobble,numbersep=5pt,escapeinside=||,numbers=left,xleftmargin=3.0ex]{C}
    --- swf_svg_add_iso_sample(read->user, ...)
    +++ sanitise(read->user, ...);
    \end{minted}
    where \mintinline[fontsize=\footnotesize]{C}{sanitise} is generated by \tool~as:
    \begin{minted}[fontsize=\footnotesize,autogobble,escapeinside=||,numbers=none,xleftmargin=0.0ex]{C}
    GF_Err sanitise(void *user, const u8 *data, ...){
      if (data == NULL) {
        return GF_BAD_PARAM;
      } else {
        return swf_svg_add_iso_sample(user,data,..);
      }  
    }
    \end{minted}
    }
\end{itemize}

These two examples highlight some issues with auto-generated code that \tool\, aims to tackle:
\begin{itemize}
    \item[-] just like seasoned developers, AI assistants may generate functionally correct but unsafe code. 
    \item[-] auto-generated code which passes the safety checks (may have latent  but not manifesting bugs) can still lead to unsafe behaviour when integrated in existing codebases. 
    \item[-] even when safety bugs are found and reported, it is not always clear how to mitigate them.
\end{itemize}

While we are yet to fully understand how developers interact with bugs in codebases with auto-generated code, we propose \tool\,to help with its safe integration into existing codebases. 

\section{Approach}
\label{sec:approach}
This section provides an overview of \tool, covers ISL background, and introduces the blame-carrying proofs. 
\subsection{Overview}

\begin{figure}[t]
\hspace{-1em}
\includegraphics[width=0.5\textwidth]{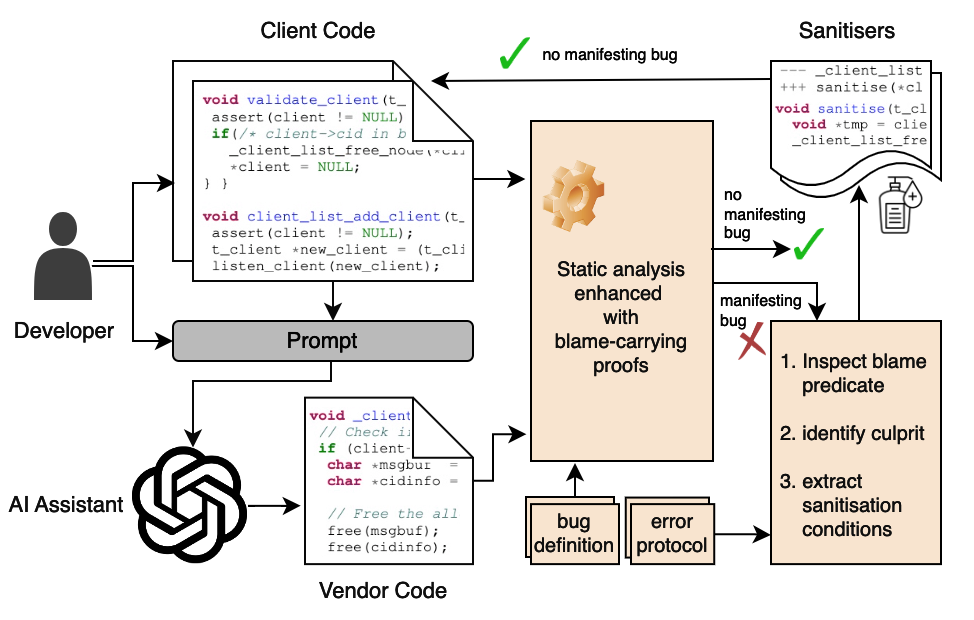}
\caption{Overview of \tool}
\label{fig:silc}
\vspace{-1em}
\end{figure}

\autoref{fig:silc} offers an overview of \tool's design. A developer working on an existing codebase (we call it client code), interacts with an AI assistant to introduce a new functionality via a carefully crafted prompt. The auto-generated code (we call it vendor code) is analysed for bugs. The analysis produces a summary describing the effect of the new functionality on the heap memory, the memory locations for which the vendor is responsible, and the program paths potentially leading to unsafe behavior. 
\tool\, next analyses the client code using the summary of the vendor code as an abstraction for the new functionality to investigate its effect on the client's memory footprint. If a manifesting bug is discovered, \tool\, inspects the bug report to determine the responsible component for the fix--the client or the vendor--and generates appropriate sanitisers to ensure the client code remains safe even in the presence of the new functionality. Note that the sanitizer is not a program repair per se, i.e. it does not fix the bug; it is a mitigation solution to avoid the unsafe path introduced during integration.






\subsection{Preliminaries}\label{sec:background}

The static analysis for bug detection is abstracting a C program using Incorrectness Separation Logic (ISL), a logic tailored to reason about the presence of memory safety bugs.

\noindent \emph{\bf Program Model.} 
The target programming language is a standard subset of C, with support for pointer manipulation operations, such as \mcode{\mallocPCword}, \mcode{\freePCword}, pointer dereference (denoted by \mcode{[x]} to avoid confusion with the separation conjunction operator), and binding (denoted by $e:=e$, where $e$ is an expression). Programs are modelled by a standard environment and state pair, 
with technical details available in \cite{Raad2020}: briefly, the environment maps the program and logical variables to values, and the state models the stack and the heap. 

\noindent \emph{\bf Assertions.}  ISL assertions (ranging over $p,q,m,f$ in \autoref{fig:logic})
have a pure and a spatial component, \mcode{k\wedge \pi}. The pure part, consisting of classical first-order logic and
Boolean assertions, is standard.
The spatial term describes the shape of the symbolic heap. Specifically, $\mcode{\emp}$ denotes an empty heap; 
$\mcode{\pointsto{x}{X}}$ describes a single memory location denoted by program pointer $\mcode{x}$ and containing the value denoted by $\mcode{{X}}$; $\mcode{\pointsto{Y}{X}}$ describes a single location denoted by logical variable $\mcode{Y}$ and containing the value denoted by $\mcode{{X}}$;
 \mcode{\dealloc{X}} indicates an invalidated location $\mcode{{X}}$; 
 \mcode{  k_1 \sep k_2} denotes a heap which can be split in two disjoint sub-heaps, one satisfying \mcode{k_1}, and one satisfying \mcode{k_2}. An ISL state of the form \mcode{(\pi;\textit{p})} also contains a program path component given as a pure term $\pi$ (required during the sanitisation generation process, \autoref{sec:automation}, \autoref{alg:pathcond}).
 
\begin{figure}[t]
  \centering
\[
{  \small
\begin{array}{l l@{\hskip 4pt}c@{\hskip 4pt}l}
 \text{Pure term} & \pi  & ::= &  \True \mid B
 \mid \pi = \pi \mid \pi \wedge \pi \mid \neg \pi 
 \\
 \text{Spatial term} &\mcode{k} & ::= &
 \mcode{\emp \mid \pointsto{x}{X}}
 \mcode{\mid \pointsto{Y}{X}} 
 \mcode{\mid \dealloc{X}}
  \mid  \mcode{  k \sep k}
 \\
 \text{Symbolic heap} & \Delta & ::= & \mcode{ k \wedge \pi}
 \\
 \text{State} & p,q,m,f& ::= & \Delta \mid \exists X.\Delta \mid (\pi;p)
 \\ [3ex]
 \multicolumn{4}{l}{
    (\text{Program vars:~} {\tt x, y})
  \quad 
    (\text{Logical vars:~}  {\tt X, Y, L, V})
  \quad
    (\text{Constants:~} {\tt c, \nil})
 }
\\
\hline
\end{array}
}
\]
\caption{ISL: Abstract domain for proving the presence of bugs. 
}
\label{fig:logic}
\vspace{-1em}
\end{figure}

\begin{figure}[t]
  \centering
\[
{  \small
\begin{array}{l}
 \okisltriple{\pointsto{x}{X}}{x:= malloc()}{\exists L. \pointsto{x}{L} \sep \pointsto{L}{V} }\\
 \okisltriple{\pointsto{x}{X}}{x:= malloc()}{\exists L. \pointsto{x}{L}  \wedge L = \nil}\\
 \okisltriple{\pointsto{x}{X} \sep \pointsto{X}{V} }{free(x)}{ \pointsto{x}{X}  \sep \dealloc{X}}\\
 \errisltriple{\pointsto{x}{X} \wedge X=\nil }{free(x)}{ \pointsto{x}{X}  \wedge X=\nil }\\
 \errisltriple{\pointsto{x}{X} \sep \dealloc{X} }{free(x)}{ \pointsto{x}{X} \sep \dealloc{X}}\\
 \okisltriple{\pointsto{x}{X} \sep \pointsto{X}{V} \sep \pointsto{y}{Y} }{[x]:=y}{ \pointsto{x}{X} \sep \pointsto{X}{Y} \sep \pointsto{y}{Y}}\\
 \okisltriple{\pointsto{x}{X} \sep \pointsto{y}{Y} \sep \pointsto{Y}{V} }{x:=[y]}{ \pointsto{x}{V} \sep \pointsto{y}{Y} \sep \pointsto{Y}{V}}\\
 \errisltriple{\pointsto{x}{X} \wedge X = \nil }{[x]:=y}{ \pointsto{x}{X} \wedge X = \nil}\\  \errisltriple{\pointsto{y}{Y} \wedge Y = \nil }{x:=[y]}{ \pointsto{y}{Y} \wedge Y = \nil }\\
 \errisltriple{\pointsto{x}{X} \sep \dealloc{X} }{[x]:=y}{ \pointsto{x}{X} \sep \dealloc{X} }\\
 \errisltriple{\pointsto{y}{Y} \sep \dealloc{Y} }{x:=[y]}{ \pointsto{y}{Y} \sep \dealloc{Y} }\\
 {[{{\color{ACMBlue} \tt  \textit{\scalebox{2}{$\ast$}}_{y_i \in pvar(e)\backslash\{x\}}  \pointsto{y_i}{Y_i} \sep
      \pointsto{x}{X} \wedge V=e[\overrightarrow{Y_i}/\overrightarrow{y_i}][X/x]}}]}\\
 \quad {\tt x:=e}\\
 {[{{\color{deepgreen} \tt \ok: \textit{\scalebox{2}{$\ast$}}_{y_i \in pvar(e)\backslash\{x\}}  \pointsto{y_i}{Y_i} \sep
      \pointsto{x}{V} \wedge V=e[\overrightarrow{Y_i}/\overrightarrow{y_i}][X/x]}}]}\\
\end{array}
}
\]
\caption{Selected predefined ISL summaries---we omit writing \mcode{\pi} when \mcode{\pi} is \mcode{\True} in \mcode{k \wedge \pi}; \mcode{pvar} returns the program variables.
}
\vspace{-1em}
\label{fig:rules}
\end{figure}

\noindent \emph{\bf Analysis.} Analysing programs with ISL involves computing summaries (specifications) for each function $\fnc$ as a set of ISL triples of the form $\isltriple{p}{\fnc}{\errlbl:q}$. This triple is said to be valid iff every state in $q$ is reachable with an exit condition $\errlbl$ (which can be $\cok$ or $\cerr$) by abstractly executing $\fnc$ on some initial state satisfying $p$.
The analysis infers specifications by using summaries of predefined instructions. Selected predefined ISL triples are given
in \autoref{fig:rules}. 
The first two summaries for \mcode{malloc}
indicate that this function can only be safely called from a program state (called pre-state) where the local variable \mcode{x} is a pointer to a location \mcode{X}. Upon executing it, the program can be in either one of the two states (called post-states): \mcode{x} points to a new memory location \mcode{L} storing a fresh value \mcode{V}, or \mcode{x} points to a new memory location \mcode{L} which is \mcode{\nil}. The latter is generally used to denote that the allocation did not succeed; \mcode{\nil} is a logical abstraction for \mcode{NULL}. Depending on the pre-state, a call to
\mcode{free} can either end up in a safe post-state where \mcode{X} is marked as having been deallocated, or in an unsafe post-state if \mcode{X} is an invalid memory location, either \mcode{NULL} or already deallocated (i.e., \mcode{X=\nil} or \mcode{\dealloc{X}}, respectively). Similarly, dereferencing succeeds safely only when the pointer being dereferenced points to a valid memory location.

At the core of the specification inference mechanisms lies 
an entailment relation between symbolic heaps and a biabductive algorithm.
An entailment $\islentails{p}{q}$ asserts that any heap satisfying $q$ also satisfies $p$. The biabuction algorithm finds the pair $(m,f)$ which makes the entailment $\islentails{p \sep m}{q \sep f}$ valid for given $p$ and $q$. 
Intuitively, $m$ is a sub-heap in $q$ but not in $p$ (nor $f$);
$f$ is a sub-heap in $p$ but not in $q$ (nor $m$).
After inferring $f$ and $m$, $m$ is being back-propagated as a \emph{missing resource} to the previous pre-states, finally being added as a required resource to the pre-condition.
$f$ is carried forward as \emph{frame} (sub-heap not referenced by the current instruction) to the post-state, finally contributing towards a stronger post-condition. 
Formally, this is achieved by the evaluation function $\mcode{Eval(\textit{p},\fnc,T)}$\footnote{For lack of space, we only discuss the \mcode{Eval} function for method call since it suffices for our extension. The rest of the rules can be found in \cite{Le2022}.}. Given a prestate $p$ and a set \mcode{T} of predefined ISL triples, \mcode{Eval} computes a set of tuples $(\errlbl, m, q)$ such that \emph{if we extend $p$ with the missing resource $m$, then $\errlbl:q$ is a valid result of executing a function \mcode{\fnc} in the state $p*m$.}

To illustrate how this works, we consider a simple function \mcode{set} that stores a value in memory via one of its parameters.
The analysis starts in a state $p$ described by \mcode{\pointsto{x}{X} \sep \pointsto{v}{V}}. Since there is no other heap information, all three summaries for the store operation  defined in \autoref{fig:rules} are used in the evaluation of \mcode{set} (we omit the details of the third store summary evaluation since it is similar to the erroneous scenario below):

\begin{minipage}[t]{0.55\linewidth}
\begin{minted}[style=bw,fontsize=\footnotesize,autogobble,escapeinside=||,numbers=none,xleftmargin=-1.5ex]{C}
void set(int* x, int v)
{|\mcode{[\bluestate{\pointsto{x}{X} \sep \pointsto{v}{V} \textit{\colorbox{light-yellow}{\mcode{\sep\, \pointsto{X}{W} }}} }]}|  
 [x] = v;
 |\mcode{[\greenstate{\ok:\pointsto{x}{X} \sep \pointsto{v}{V} \sep \pointsto{X}{V}}]}| 
}
\end{minted}
\end{minipage}
\hfill
\begin{minipage}[t]{0.4\linewidth}
\begin{minted}[style=bw,fontsize=\footnotesize,autogobble,escapeinside=||,numbers=none,xleftmargin=-3.2ex]{C}
void set(int* x, int v)
{|\mcode{[\bluestate{\pointsto{x}{X} \sep \pointsto{v}{V} \textit{\colorbox{light-yellow}{\mcode{\wedge\, {X}=\nil }}} }]}|  
 [x] = v;
 |\mcode{[\redstate{\err:\pointsto{x}{X}\sep \pointsto{v}{V} \wedge {X} = \nil} ]}| 
}
\end{minted}
\end{minipage}

The analysis infers the missing resource $m$ (\colorbox{light-yellow}{highlighted} above) which extends $p$ into a state $p \sep m$ enabling the use of the corresponding triple. 
This missing $m$ is then backpropagated into the specification of \mcode{set} resulting in three scenarios: one safe and two erroneous specifications.  
Since the erroneous ones depend on the validity of parameter \mcode{x}, these issues are not reported to the user and are termed \emph{latent bugs}. Only if a caller invokes \mcode{set} with an invalid argument \mcode{x}, 
would these issues be reported as \emph{manifesting bugs} at the caller's side.

\subsection{Blame-Carrying ISL}\label{sec:blame-shifting-framework}

In this work, we extend the ISL logic by introducing the assertions in \autoref{fig:blameislassertions}. These assertions help to identify the responsible side for a reported safety violation, i.e. the culprit. The core of the extension is the abstraction of blame as a higher-order predicate \mcode{\cblame}, and an auxiliary predicate \mcode{\cworld} which tracks the context of the considered bugs.

For a predicate \mcode{\blame{X}{E}{\bugcond}{S}{C}} we define an entity \mcode{E} to be the side responsible for the manipulation of resource \mcode{X}. Entity \mcode{E} can be either \mcode{Client} or \mcode{Vendor} if the responsible side is known, or a logical variable \mcode{W} for an unknown entity. 
If the unsafety condition $\bugcond$, which is either an ISL assertion $p$ 
or an ISL predicate \mcode{\bug{P}{X}}, 
holds true in a program state, say $q$, that is, \mcode{\islentails{q}{\bugcond}}, then \tool\, applies the sanitisation denoted by \mcode{S}. A sanitisation \mcode{S} describes how the interaction between the client and the vendor code should be handled on a given program path $\pi$ in order to avoid the safety condition violation. $\pi$ is automatically extracted from the specification of the vendor code (the extraction is described in \autoref{sec:automation}) and it represents the logical condition (pure term) which needs to hold for the latent bug to become a manifest bug.  For example, given a bug condition \mcode{P(X) := {\exists x. \pointsto{x}{X} \wedge {X} = \nil}} and erroneous state $q$ such that \mcode{\islentails{\textit{q}}{P(X)}},
a blame-predicate instance  {\small \mcode{\blame{X}{\tclient}{P(X)}{\sanitise{({X} = \nil)}{stop}}{C}}} states that the responsible side for the erroneous condition \mcode{P(X)} on manipulating the resource abstracted by \mcode{X} in state $q$ is \mcode{\tclient}. To avoid this bug \tool\, applies a sanitisation which \mcode{stop}s the program's execution on the program path where \mcode{X = \nil}. The program path for sanitisation is given as a logical formula, using logical variables. Translating this logical condition into one on program variables is described in \autoref{sec:automation}.

\begin{figure}[t]
\centering
 \begin{subfigure}[b]{0.5\textwidth}
\[
{  \small
\begin{array}{l l c l}
 \text{Spatial term} &\mcode{k} & ::= &
 \ldots \mid \mcode{\blame{X}{E}{\bugcond}{S}{C}} \mid \mcode{\world{E}{P}{S}{C}}
 \\
 \text{Entity} & \mcode{E} & ::= & \tclient \mid \tvendor \mid \mcode{W}
 \\
 \text{Bug~definition} & \mcode{bdef} & ::= & \mcode{\bug{P}{X}} := p
 \\
 \text{Bug~condition} & \mcode{\bugcond} & ::= & \mcode{\bug{P}{X}}  \mid p
 \\
 %
\text{Sanitisation} & \mcode{S} & ::= & \mcode{\sanitise{\pi}{stop}} \mid \mcode{\sanitise{\pi}{noLeak}}
 \\ [3ex]
 \multicolumn{2}{l}{
    (\text{Bug reference:} \quad {\tt P})}
 &
    \multicolumn{2}{r}{    
    (\text{Program code template:} \quad \mcode{stop,~ noLeak})
}
\\
\hline
\end{array}
}
\]
\caption{Blame-carrying ISL.}
\label{fig:blameislassertions}
\end{subfigure}
 \begin{subfigure}[b]{0.5\textwidth}
 \vspace{1em}
\[
{  \small
\begin{array}{l l c l}
 \text{Bug kind} & \mcode{b} & ::= & \mcode{NPD} \mid \mcode{MemLeak}  \mid \mcode{UAF} \\
 \text{Bug definition} & \multicolumn{3}{l}{\mcode{\bugdef{b}{bdef}}}\\
 \text{Err protocol} & \multicolumn{3}{l}{\mcode{\errprot{b}{stop}} \mid \mcode{\errprot{b}{noLeak}}}
\end{array}
}
\]
\caption{User inputs: bug definition and sanitisation policy.}
\label{fig:userinputs}
\end{subfigure}
\caption{Components of \tool. }
\label{fig:silccomponents}
\vspace{-1em}
\end{figure}

\begin{figure*}[t]
  \centering
\[
{
\hspace{-1ex}
\small
\begin{array}{l}
 \okisltriple{\pointsto{x}{X} \sep \world{E}{P}{S}{C}}{x:= malloc()}{\exists L. \pointsto{x}{L} \sep \pointsto{L}{V} \sep \blame{L}{E}{\bug{P}{L}}{S}{C}  \sep \blame{V}{E}{\bug{P}{V}}{S}{C}}\\
 \okisltriple{\pointsto{x}{X}\sep \world{E}{P}{S}{C}}{x:= malloc()}{\exists L. \pointsto{x}{L} \sep  \blame{L}{E}{\bug{P}{L}}{S}{C} \wedge L = \nil}\\
 \okisltriple{\pointsto{x}{X} \sep \pointsto{X}{V} \sep \world{E}{P}{S}{C} \sep \blame{X}{\_}{\bug{P}{X}}{\_}{\_} }{free(x)}{ \pointsto{x}{X}  \sep \dealloc{X} \sep \blame{X}{E}{\bug{P}{X}}{S}{C} }\\
 \okisltriple{\pointsto{x}{X} \sep \pointsto{X}{V} \sep \pointsto{y}{Y} \sep \world{E}{P}{S}{C} \sep \blame{X}{\_}{\bug{P}{X}}{\_}{\_}  }{[x]:=y}{ \pointsto{x}{X} \sep \pointsto{X}{Y} \sep \pointsto{y}{Y} \sep \blame{X}{E}{\bug{P}{X}}{S}{C}}\\
 \okisltriple{\pointsto{x}{X} \sep \pointsto{y}{Y} \sep \pointsto{Y}{V} \sep \world{E}{P}{S}{C} \sep \blame{Y}{\_}{\bug{P}{Y}}{\_}{\_}  }{x:=[y]}{ \pointsto{x}{V} \sep \pointsto{y}{Y} \sep \pointsto{Y}{V} \sep \blame{Y}{E}{\bug{P}{Y}}{S}{C}}\\
 {[{{\color{ACMBlue} \tt  \textit{\scalebox{2}{$\ast$}}_{y_i \in pvar(e)\backslash\{x\}}  \pointsto{y_i}{Y_i} \sep
      \pointsto{x}{X}  \sep \world{E}{P}{S}{C} \wedge V=e[\overrightarrow{Y_i}/\overrightarrow{y_i}][X/x]}}]}
 {\tt\, x:=e \,}
 [{{\color{deepgreen} \tt \ok: \textit{\scalebox{2}{$\ast$}}_{y_i \in pvar(e)\backslash\{x\}}  \pointsto{y_i}{Y_i} \sep
      \pointsto{x}{V}  \sep \blame{V}{E}{\bug{P}{V}}{S}{C} }}
 \\
 \multicolumn{1}{r}{{{\color{deepgreen} \tt  \wedge V=e[\overrightarrow{Y_i}/\overrightarrow{y_i}][X/x]}}]}\\
\end{array}
}
\]
\vspace{-3ex}
\caption{ Blame-carrying ISL summaries (erroneous triples follow those in \autoref{fig:rules}).}
\label{fig:blameisl}
\vspace{-1em}
\end{figure*}

A predicate \mcode{\world{E}{P}{S}{C}} is used to denote the world, that is, the code, of side \mcode{E}\footnote{An entity \mcode{E} also stores information such as filename, function name and line number to indicate blame. For brevity, we omit these details from the formalisation since they are straightforward to deal with.} for a bug condition referred to by \mcode{P} with corresponding sanitisation protocol described by \mcode{S}. The blame may shift from one entity to another according to the world the analysed code belongs to. 
Let us see how these predicates play together for assigning blame by revisiting the \mcode{set} function introduced in the previous subsection. 

\begin{minipage}[t]{0.55\linewidth}
\begin{minted}[style=bw,fontsize=\footnotesize,autogobble,escapeinside=||,numbers=none,xleftmargin=-1.5ex]{C}
void set(int* x, int v)
{|\mcode{[\bluestate{ \pointsto{x}{X} \sep \ldots \sep \world{\boxed{\tvendor}}{\_}{\_}{} \textit{\colorbox{light-yellow}{\mcode{\sep\, \blame{X}{\boxed{\tt W}}{\_}{\_} }}} }]}| 
 [x] = v;
 |\mcode{[\greenstate{\ok: \pointsto{x}{X} \sep \ldots \sep \world{\tvendor}{\_}{\_}{}  \sep\, \blame{X}{\boxed{\tvendor}}{\_}{\_}   }]}| 
}
\end{minted}
\end{minipage}

We ignore for now the bug condition and the sanitisation attributes, and just focus on assigning responsibility for the use of resource \mcode{X}, pointed to by \mcode{x}.
The \mcode{\cworld} predicate indicates that the analysed code is auto-generated code as it resides in the \mcode{\tvendor} world. Before the dereference of \mcode{x} there was an unknown entity \mcode{W} which was responsible for the manipulation of resource \mcode{X}. However, after the successful dereference the \mcode{\tvendor} becomes the responsible entity. The intuition behind this shift is simple and it follows the so-called ``dish-washing principle": whichever entity touches/accesses the resource last becomes responsible for it.
To note that the \mcode{\colorbox{light-yellow}{\small \tblame}} predicate instance in the prestate is inferred as a missing resource by the \mcode{Eval} function. \mcode{Eval} makes use of the following modified predefined ISL triple for the store operation when evaluating \mcode{set}: \\
\[\small
\hspace{-1em}
\begin{array}{cc}
\left[ \bluestate{\mcode{\pointsto{x}{X} \sep \pointsto{X}{V} \sep \pointsto{y}{Y} \sep \world{\boxed{\tt E}\tikzmark{start1}}{\_}{\_}{C} \sep \blame{X}{\boxed{\tt W}}{\_}{\_}{\_}  }}\right]\\
{[x]:=y}\\
\left[\greenstate{\ok:\mcode{\pointsto{x}{X} \sep \pointsto{X}{Y} \sep \pointsto{y}{Y} \sep \world{E}{\_}{\_}{C} \sep \blame{X}{\tikzmark{stop1}\boxed{\tt E}}{\_}{\_}{C}}}\right]
\end{array}
\begin{tikzpicture}[remember picture]
 \draw [overlay, ->, line width=0.5pt, black] (pic cs:start1)  ++ (0.1,-0.1) --  ( $ (pic cs:stop1) +(-0.1,+0.33) $ ) ;
\end{tikzpicture}
\]

This triple indicates that after a store operation, the blame for \mcode{X} shifts  from the unknown entity \mcode{W} to entity \mcode{E} representing the current world. Note that \mcode{W} could be \mcode{E}, in which case the responsible entity remains the same. Treating the blame and world like ISL assertions has the benefit 
of supporting blame proofs  without changing the biabductive core reasoning.
The blame predicate required for this reasoning are seamlessly inferred by the \mcode{Eval} function described in \autoref{sec:background}, resulting in the specification of \mcode{set} described earlier. 

With this seamless integration of blame, the predefined erroneous triples become pleasantly simple, and they do not need to assert blame explicitly. For example, the triple for the NPD of the store operation is modified as follows:
\[\small
\hspace{-1em}
\begin{array}{cc}
\left[ \bluestate{\mcode{\pointsto{x}{X} \sep \world{\tikzmark{start2}\boxed{\tt E}}{\_}{\_}{C} \wedge X= \nil  }}\right]\\
{[x]:=y}\\
\left[\redstate{\err^{\boxed{\tt \scriptscriptstyle E}}\tikzmark{stop2}:\mcode{\pointsto{x}{X}  \sep \world{E}{\_}{\_}{C}  \wedge X= \nil  }}\right]
\end{array}
\begin{tikzpicture}[remember picture]
 \draw [overlay, ->, line width=0.5pt, black] (pic cs:start2)  ++ (-0.05,-0.1) --  ( $ (pic cs:stop2) +(+0.05,+0.37) $ ) ;
\end{tikzpicture}
\]

There are a few subtle observations about the triples with blame. First, when an operation ``fails" for safety reasons, there is no blame shift; the responsibility for the resource remains with the last entity that successfully accessed it. 
Formally, the blame predicate for resource \mcode{X} is framed and thus preserved, regardless of the state $p$ in which this triple is evaluated. This applies to all erroneous triples, leading to a second observation: an erroneous state retains its blame predicates regardless of the calling context. In other words, the specifications remain highly composable while maintaining precise blame information.  
This is what allows \tool\, to \emph{identify the side responsible} for the resource leak in \autoref{fig:exmemleak}.
The last observation refers to the superscript of \mcode{\err^{\tt E}} to indicate the world \mcode{E} in which this latent bug would manifest if triggered.
This is what allows \tool\, to \emph{precisely find the manifestation} of memory safety bugs even in the presence of complex call chains as the one in \autoref{fig:npd}.

Since a blame analysis may tackle multiple kinds of bugs involving the same resource, the world and the blame predicates must distinguish between these different bugs. That is the reason why the blame predicate has an ISL condition \mcode{\bug{P}{X}} corresponding to the bug reference \mcode{P} the world is defined for. For instance, for an assertion  
{\small \mcode{\pointsto{x}{X} \sep \world{E}{P}{\_}{C} \sep \blame{X}{E}{\bugcond}{\_}{\_}}} where {\small \mcode{P(Z) := {\exists z. \pointsto{z}{Z} \wedge {Z} = \nil}}}, the world and blame predicates are related iff \mcode{\islentails{\beta}{\bug{P}{X}}}. 

The sanitization template specifies how a particular bug, if triggered, should be managed.
Although including a template as a parameter for the two special predicates seems at odds with a logic-based bug detection solution, this approach allows \tool\, to customize the sanitisation according to both the type of bug and the context in which it was detected.
A more complete predefined ISL triple with blame for the store operation looks as follows:  
\[\small
\hspace{-1em}
\begin{array}{cc}
\left[ \bluestate{\mcode{\pointsto{x}{X} \sep \ldots \sep \world{E}{P}{\boxed{\tt S}\tikzmark{start3}}{C} \sep \blame{X}{W}{\bug{P}{X}}{\boxed{\tt S'}}{\_}  }}\right]\\
{[x]:=y}\\
\left[\greenstate{\ok:\mcode{\pointsto{x}{X} \sep \ldots \sep \world{E}{P}{S}{C} \sep \blame{X}{E}{\bug{P}{X}}{\tikzmark{stop3}\boxed{\tt S}}{C}}}\right]
\end{array}
\begin{tikzpicture}[remember picture]
 \draw [overlay, ->, line width=0.5pt, black] (pic cs:start3)  ++ (0.1,-0.1) --  ( $ (pic cs:stop3) +(-0.1,+0.33) $ ) ;
\end{tikzpicture}
\]

As the blame for resource \mcode{X} shifts from \mcode{W} to \mcode{E}, so does the sanitisation policy. For example, given an NPD bug, the sanitisation policy \mcode{S'} coming from the world of \mcode{W} could indicate that the program execution should terminate with
\mintinline[fontsize=\footnotesize,escapeinside=||]{C}{return;}
on the buggy path, while in the world of \mcode{E}, the sanitisation \mcode{S} could use  
\mintinline[fontsize=\footnotesize,escapeinside=||]{C}{goto exit;} to indicate program termination.

\autoref{fig:blameisl} introduces the predefined ISL triples modified to include blame. We treat the \mcode{\cworld} predicate as a persistant resource, i.e. one which once defined it does not change on the course of the evaluation process. What that means for our rules is that we add it in the prestate of the triple, but, for brevity, we omit adding it in the poststate. We use $\_$ to denote an anonymous value. 
The rules for \mcode{malloc} introduce new blame predicates, while the rules for \mcode{free} and dereference indicate blame shifting according to the world predicate. We omit detailing the erroneous rules since they are identical with those in \autoref{fig:rules} with the exception of the superscript of  \mcode{\err^E} to indicate the world of error manifestation: $\errblameisltriple{p \sep\world{E}{\_}{\_}{}}{\fnc}{q}{E}$. 

The triples indicate that the blame is shifted between the worlds along with the culprit resource. However, there is one exception. Once \mcode{Vendor} assumes responsibility for a resource, this responsibility remains with \mcode{Vendor}, even if it makes calls to \mcode{Client} functions within its body, i.e. it does not flow into \mcode{Client}. This design follows the intuition that a developer might have zero trust in \mcode{Vendor}'s execution; therefore, it is \mcode{Vendor}'s responsibility to address the unsafety caused by calls to existing \mcode{Client} functions. We enforce this rule through the definition of the \mcode{Eval} function for function call, ensuring that the inferred \mcode{Blame} predicates inherit the current \mcode{World}'s attributes when the world pertains to \mcode{Vendor}. 

It may seem like the code requires many annotations. However, most are automatically derived from the bug definitions and error protocol provided by the user according to \autoref{fig:userinputs}. We show how this process is automatized in \autoref{sec:automation}. 

\subsection{Automation}\label{sec:automation}
A user of \tool\, only provides the information in \autoref{fig:userinputs}; everything else is derived automatically. 
The user states what kind of bugs the blame-shifting framework tackles, 
what ISL assertions describe them (\mcode{bdef}, extracted from the bug detection definitions used by \pulse), and what should the sanitisation policy be. 
This information is supplied once per project, after which the analysis automatically instantiates it for each function according to \autoref{alg:auto}.
Given a function \mcode{fnc}, the analysis creates a world predicate for each considered bug type (line 3). A conjunction of these world predicate instances serves as a precondition to kick-start the analysis (line 4). Once a manifesting bug is identified (line 5), \tool\, proceeds with inspecting the state in which it failed. If the error occurs in a different world than the one responsible for the unsafe resource, then the manifesting bug is a result of the integration (line 7). In this case \tool\, proceeds by sanitizing the call to the auto-generated function, applying the ascribed sanitization policy (line 10) after automatically extracting the path conditions that need to be disabled (line 8). 

\begin{algorithm}[t]
	\small
	\caption{\algo{Analysis}}
	\label{alg:auto}
    \SetKwInOut{Input}{Input}\SetKwInOut{Output}{Output}
	\Input{A function \mcode{fnc}, a set \mcode{B} of considered bugs, a map \mcode{\mathcal{B}} of bug definitions, a map \mcode{\mathcal{S}} of error protocols }
	\Output{\mcode{fnc'} - the sanitised \mcode{{fnc}} }
    \mcode{E \leftarrow} (\mcode{ fnc} is \mintinline[fontsize=\footnotesize,escapeinside=||]{C}{|@|Vendor} annotated) ? \mcode{ \tvendor:\tclient}\\
    $p \leftarrow $ conjunction of \mcode{\world{E}{\mathcal{B}(b)}{\mathcal{S}(b)}{}} for each \mcode{b\in B}\\
    \mcode{T \leftarrow} set of blame-carrying ISL triples (\autoref{fig:blameisl})\\
    ${(\errlbl,m,q) \leftarrow}$ \mcode{Eval(\textit{p},fnc.body,T)}\tcc*{analyse \mcode{fnc}}
    \If{\mcode{\errlbl} is \mcode{\redstate{\err^{E}}}}{
    \mcode{\blame{\_}{E'}{\_}{S}{} \leftarrow} extract blame predicate from $m$\\
    \If(\tcc*[f]{integration bug}){\mcode{E\neq E'}}{ 
    \mcode{\pi, call \leftarrow} \algo{ExtractPathCondition} ($p\sep m$, $\errlbl:q$)\\
    \mcode{S'\leftarrow} insert $\pi$ into \mcode{S} \\
    \mcode{fnc' \leftarrow} sanitise the \mcode{call} to vendor function with \mcode{S'}
    
    }
    }
\end{algorithm}      

A bug may manifest due to resources flowing into the vendor code (+) or out of it (-). The direction of this flow determines where the sanitisation occurs, either before or after the call to the vendor function. The NPD in  \autoref{fig:npd} occurs because of an invalid argument passed on to the vendor code, hence its sanitisation is applied before the call to the vendor code. The memory leak in  \autoref{fig:exmemleak}  occurs due to memory being leaked out of the vendor code, hence the  sanitisation is applied after the call to the vendor code.  
To indicate the flow direction, we add a sign attribute to the sanitisation template: \mcode{S^{+/-}}. 

\begin{algorithm}[t]
	\small
\caption{\algo{ExtractPathCondition}}
	\label{alg:pathcond}
    \SetKwInOut{Input}{Input}\SetKwInOut{Output}{Output}
	\Input{ manifest error pre-/post-states \mcode{\errblameisltriple{p\sep \blame{X}{E}{\_}{S}{}}{,}{q}{E'}}}
	\Output{vendor \mcode{call} and erroneous path \mcode{\pi} to be sanitised}
   \If{\mcode{S} has \mcode{``+"} attribute}{ 
       \mcode{call \leftarrow } retrieve the vendor call introducing \mcode{\redstate{\err^{E'}}}\\
       \mcode{fnc \leftarrow } retrieve the function corresponding to \mcode{call}\\
       $p'$ \mcode{\leftarrow} precondition of \mcode{fnc} which enables  \mcode{\redstate{\err^{E'}}}\\ 
       \mcode{\pi' \leftarrow } remove the spatial assertions from $p'$\\
       \mcode{\pi_0 \leftarrow } substitute \mcode{fnc} params in $\pi'$ with its \mcode{call} args}
   \If{{\mcode{S} has \mcode{``-"} attribute}}{
      \mcode{call \leftarrow } retrieve the vendor call introducing \mcode{E}\\
      \mcode{fnc \leftarrow } retrieve the function corresponding to \mcode{call}\\
      \mcode{q' \leftarrow } postcond of \mcode{fnc} introducing \mcode{\blame{X}{E}{\_}{S}{}}\\
      \mcode{\pi' \leftarrow} extract path condition from vendor's $q'$\\
      \mcode{\pi_0' \leftarrow} substitute \mcode{fnc} params in $\pi'$ with its \mcode{call} args\\
      \mcode{\pi'' \leftarrow} extract path condition from client's $p$ \\
      \mcode{\pi_0 \leftarrow \pi_0' \wedge \pi''} 
      }
   \mcode{\pi \leftarrow } collapse logical assertions in \mcode{\pi_0} to program vars
\end{algorithm}      

\tool\, automatically computes the program path condition on which the sanitisation needs to intervene,
This process is described in \autoref{alg:pathcond}. For a given manifestation error 
described by an erroneous ISL triple, \mcode{\errblameisltriple{p \sep \blame{X}{E}{\_}{S}{}}{inst}{q}{E'}}, if sanitisation policy \mcode{S} indicates an issue with the resources flowing into the vendor code (line 1) then the manifesting bug lies within the vendor function. \tool\, retrieves this function using information stored in \mcode{E'} (lines 2-3), and retrieves the ISL triple which introduces \mcode{\redstate{\err^{E'}}} (line 4). The precondition of this triple specifies the condition triggering this error. However, the precondition contains spatial terms that need to be transformed to program constructs. \tool\, removes spatial terms (line 5) by approximating assertions of the form \mcode{\pointsto{x}{X} \sep \pointsto{X}{V}} with \mcode{NULL} checks of the form \mcode{x \neq \nil}.
This conservative approximation ensures that the manifesting error is disabled 
by the sanitisation.
If sanitisation policy \mcode{S} indicates an issue with a resource flowing out of vendor code (line 7) then the manifesting bug lies within the client function. The blame predicate \mcode{\blame{X}{E}{\_}{S}{}} identifies the vendor function that last accessed resource \mcode{X} (lines 8-9).
The path condition which disables this bug is a conjunction of the vendor's path condition on which the culprit resource was last accessed (extracted from the corresponding postcondition, line 10) and the client's path condition on which the bug manifests (lines 11-14). 
\tool\, finally translates the path condition from first order logic formula to a program condition that only quantifies over program variables and program constructs (line 15). For example, in an abstract program state where \mcode{\pointsto{x}{X}\sep\pointsto{X}{Y} \sep \pointsto{Y}{V}}, a first order logical condition \mcode{Y\neq nil}  is translated into a program condition  \mintinline[fontsize=\footnotesize,escapeinside=||]{C}{*x != NULL}.
Since these are straightforward fixed translation rules, we will omit the details due to space constraints.

\section{Evaluation}
\label{sec:evaluation}
In this section we assess the safety of the integration of auto-generated code, and evaluate the effectiveness of \tool. Specifically, we aim to answer the following questions:
\begin{itemize}[leftmargin=+.5in]
    \item[\bf RQ1] Is LLM producing safer code than developers did for codebases with historically known issues?
    \item[\bf RQ2] Can LLM introduce unsafety in code with no previously known safety issues?
    \item[\bf RQ3] How effective is the blame shifting framework at aligning auto-generated code with existing code? 
\end{itemize}

\subsection{Setup}
\tool\,consists of about 9k lines of Python, Rust and OCaml code, including glue code for existing tools and libraries, such as \pulse, csnake, openAI API, and Z3. 

\noindent \emph{\bf Prompt.} The prompt format used by \tool\, follows the standards of the prompt engineering practices used for the evaluation of AI models for code generation \cite{HumanEval,MBPP,EvoCodeBench}. The prompt has two sections. A static one which is common for all tasks and where the AI assistant is endowed with an expertise and is given its goal and guidelines:

\begin{boxprompt}
 |\emph{You are a coding assistant performing code completion}|
 |\emph{tasks. You will generate a function from comments and }|
 |\emph{function signature. As input, you will be given 4 parts:}| 
 
 1. Code **before** the function to be generated, in 
     format <prefix> PREFIX_CONTENT </prefix>
 2. Code **after** the function to be generated, in 
     format <suffix> SUFFIX_CONTENT </suffix>
 3. Comment describing the function to be generated, in 
     format <comment> COMMENT_CONTENT </comment>
 4. Header of the function to be generated, in 
     format <header> HEADER_CONTENT </header>
     
 |\emph{As output, you should generate a complete function, }|
 |\emph{including the function header and body.}|
\end{boxprompt}

The second section is unique to each task and it follows the format provided
in the static section. It starts by sharing the context of the vendor function (about 100 lines before and after the call), it next describes what the function is supposed to do (according to existing comments) and shares its header.

\noindent \emph{\bf Dataset.} We collected 28 
open-source C projects that are relatively large ($\ge$ 6k LOC), manipulate pointers, are actively maintained,
 popular (40 to 25k GitHub stars), and have a build command  compatible with  \pulse.  We excluded 9 projects with no known cybersecurity vulnerabilities\cite{cve}
keeping those with past reports of Null Pointer Dereference (NPD), Memory
\begin{wrapfigure}[7]{r}{0.35\linewidth}
\vspace{-1em}
    \begin{tikzpicture}[scale=1.1]
    \hspace{-0.8em}
    \newcounter{a}
    \newcounter{b}
    \foreach \p/\t/\c in {37/NPD/NPDunsafecolour, 49/ML/MLunsafecolour, 14/UAF/UAFunsafecolour}
      {
        \setcounter{a}{\value{b}}
        \addtocounter{b}{\p}
        \slice{\thea/100*360}
              {\theb/100*360}
              {\p\%}{\t}{\c}
      }
    \end{tikzpicture}
\end{wrapfigure}
 Leak (ML), and Use-After-Free (UAF). From 
the remaining 19 projects, we selected 49 historically unsafe 
functions (we call this the CVE dataset; with distribution of bug types as illustrated in the chart on the right) and 53 randomly selected functions that manipulate pointers and, historically, have no prior safety issues (we refer to it as the unbiased dataset).


\noindent \emph{\bf Configuration.} The evaluation queries both the popular \gptthree-turbo (gpt-3.5-turbo-0125) and the latest \gptfour (gpt-4o-2024-05-13). All experiments are conducted on Ubuntu 20.04.6 LTS system on a desktop computer with a 24-Core 3.0 GHz Intel processor and 64GB of RAM.

\subsection{RQ1 - does LLM generate safer code than developers did?}
We first assess the safety of code generated by \gpt\, compared to developer-written code for the 49 functions with historically known safety issues. 
Functional correctness is not evaluated, as our focus is solely on the safety of the auto-generated code.
For each subject, we removed the unsafe function and asked \gpt\, to regenerate it given the function's calling context, header, and functionality description as prompt. Each version of \gpt\, generated 5 variants per subject, totaling 490 functions, which we validated for safety both manually and using \pulse. The manual confirmation of the bugs was done by comparing the manifesting bugs against the corresponding publicly available CVE reports.

\begin{figure}[t]
\hspace{0.5em}
\begin{subfigure}{0.45\linewidth}
    \begin{tikzpicture}[scale=1.1]
    \foreach \pp/\tt/\cc/\ppp/\ttt/\ccc in {11//NPDunsafecolour/26/NPD/NPDsafecolour, 21//MLunsafecolour/28/ML/MLsafecolour,6//UAFunsafecolour/8/UAF/UAFsafecolour}
      {
        \setcounter{a}{\value{b}}
        \addtocounter{b}{\pp}
        \slicenolbl{\thea/100*360}
              {\theb/100*360}
              {\pp\%}{\tt}{\cc}
        \setcounter{a}{\value{b}}
        \addtocounter{b}{\ppp}
        \slicefull{\thea/100*360}
              {\theb/100*360}
              {\ppp\%}{\ttt}{\ccc}
      }
    \newcounter{aa}
    \newcounter{bb}
    \foreach \pp in {37,49,14}
      {
        \setcounter{aa}{\value{bb}}
        \addtocounter{bb}{\pp}
        \sliceline{\theaa/100*360}
              {\thebb/100*360}
      }
    \newcounter{aaa}
    \newcounter{bbb}
    \foreach \pp in {11,26,21,28,6,8}
      {
        \setcounter{aaa}{\value{bbb}}
        \addtocounter{bbb}{\pp}
        \innerlabel{\theaaa/100*360}
              {\thebbb/100*360}
              {\pp\%}
      }
    \end{tikzpicture}
    \caption{CVE dataset (\autoref{tab:CVE})}\label{chart:CVE}
\end{subfigure}
\hfill
\begin{subfigure}{0.45\linewidth}
    \centering
    \begin{tikzpicture}[scale=1.1]
    \foreach \p/\t/\c in {34/NPD/NPDunsafecolour, 66/ML/MLunsafecolour}
      {
        \setcounter{a}{\value{b}}
        \addtocounter{b}{\p}
        \slice{\thea/100*360}
              {\theb/100*360}
              {\p\%}{\t}{\c}
      }
    \end{tikzpicture}
    \caption{Unbiased dataset (\autoref{tab:autogen})}\label{chart:auto-gen}
\end{subfigure}
\caption{Manifest bug type distribution per dataset}\label{fig:bugdistribution}
\vspace{-1em}
\end{figure}
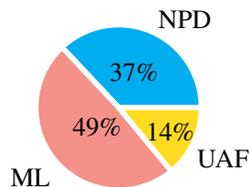
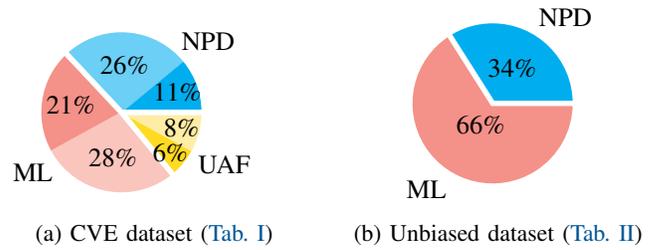

\noindent\emph{\bf Results}. \autoref{tab:CVE} summarises the evaluation results. Of the 490 auto-generated functions, only 86 generated by \gptthree\, were syntactically valid (i.e. non-empty and compile successfully), and 84 for \gptfour. Furthermore, \gpt\, generated safe functions for 29 subjects (see the \emph{Manifest Bug} column). However, the amount of bugs introduced for the other 10 subjects seems to indicate a cumulative bug rate of over 50\% for both {\gpt}s. That is, more that 50\% of the syntactically valid auto-generated functions reproduce the original bug or introduce new ones. The distribution of manifesting bug types is illustrated in \autoref{chart:CVE}, where the darker shades in each slice indicate the percentage of functions reproducing the original bug, while the lighter ones indicate auto-generated functions which are safe. Although only in a few projects, new latent bugs could pose concerns for an evolving project.

\begin{tcolorbox}[blanker,top=3mm,bottom=3mm,borderline horizontal={2pt}{0pt}{gray}]
\emph{Overall, \gpt-generated code is safer than developer-written code for reported unsafe functions in real-world projects, but it still has a significant unsafety rate of over 50\%.}
\end{tcolorbox}

\begin{table*}[htbp]
    \footnotesize
    \centering
    \caption{\small CVE Dataset - statistics for experiments on 19 projects with sizes in thousands of lines of code (kLoC) comprising 49 subjects (\emph{\#Fnc}) with an average size in lines of code (LoC). The \emph{Syntactically Valid} column indicates the total number of compilable and non-empty auto-generated functions out of the total of \emph{\#Fnc}$\times 5$ variants for \gptthree\, and \gptfour, respectively. The \mcode{m}+\mcode{n} under the \emph{Manifest Bugs} indicates that m autogenerated functions reproduce the known manifest bug when integrated into the corresponding project, and that n autogenerated functions introduce new manifesting bugs. The \emph{Latent Bug} column indicates the number of new, non-manifesting bugs. \emph{Sanitises} indicates how many of the \emph{Manifest Bugs} \tool\, successfully sanitised. \emph{Average Time} is the time per project needed to analyse and sanitise one auto-generated function (\emph{Blame} is the overhead for supporting blame, and \emph{San} is the time required to generated a sanitizer).}\label{tab:CVE}
    \begin{tabularx}{\textwidth}{lYY*{2}{Y}*{2}{Y}*{2}{Y}*{2}{Y}*{2}{Y}*{2}{Y}*{2}{Y}}
        \toprule
        \multirow{2}{*}{\emph{Project}} & \multirow{1}{*}{\emph{kLoC}} & \multirow{1}{*}{\emph{\#Fnc}} & \multirow{1}{*}{\emph{Avg.}} & \multicolumn{2}{c}{\emph{Syntactically Valid}} &\multicolumn{2}{c}{\emph{Manifest Bugs}} & \multicolumn{2}{c}{\emph{Latent Bugs}} & \multicolumn{2}{c}{\emph{Sanitised}} & \multicolumn{3}{c}{\emph{Average Time(s)}} \\
        \cmidrule(l){5-6}  \cmidrule(l){7-8} \cmidrule(l){9-10} \cmidrule(l){11-12} \cmidrule(l){13-15}
        &  & \multirow{1}{*}{}& \multirow{1}{*}{\emph{LoC}}&  3.5 & 4o & 3.5 & 4o & 3.5 & 4o & 3.5 & 4o & \emph{Total} & \emph{Blame} & \emph{San} \\
        \midrule
        libming & 53.6 & 5 & 27 & 11 & 9 & \oldnew{6}{2} & \oldnew{5}{5} & 10 & 25 & \saneval{100} & \saneval{100} & 285 & 1.38 & 0.12\\ 
        ovs & 280 & 1 & 51 & 5 & 4 & \oldnew{0}{0} & \oldnew{0}{0} & 0 & 0 & \na & \na & 85 & 14.9 & 0.07\\ 
        OpenNDS & 6.3 & 1 & 23 & 5 & 5 & \oldnew{5}{5} & \oldnew{5}{5} & 0 & 0 & \saneval{100} & \saneval{100} & 4.8 & 0.98 & 0.43\\ 
        matio & 19.9 & 1 & 36 & 1 & 4 & \oldnew{0}{0} & \oldnew{1}{1} & 0 & 9 & \na & \saneval{100} & 23 & 7.96 & 0.06\\ 
        atheme & 83.8 & 1 & 38 & 3 & 5 & \oldnew{0}{0} & \oldnew{0}{0} & 43 & 48 & \na & \na & 84 & 55.1 & 0.02 \\ 
        freeglut & 27.0 & 2 & 21 & 10 & 10 & \oldnew{10}{0} & \oldnew{6}{0} & 0 & 2 & \saneval{100} & \saneval{100} & 60 & 36.9 & 0.21\\ 
        gpac & 690 & 16 & 205 & 8 & 11 & \oldnew{2}{5} & \oldnew{3}{3} & 0 & 7 & \saneval{14} & \saneval{83} & 726 & 65.2 & 1.22\\ 
        lotos & 25.1 & 1 & 14 & 5 & 3 & \oldnew{1}{4} & \oldnew{0}{0} & 0 & 0 & \saneval{100} & \na & 2.4 & 0.12 & 0.60\\ 
        media-server & 42.8 & 2 & 47 & 5 & 5 & \oldnew{0}{0} & \oldnew{0}{0} & 0 & 0 & \na & \na & 225 & 44.2 & 0.36\\ 
        ntfs-3g & 86.5 & 1 & 55 & 0 & 2 & \na & \oldnew{0}{0} & \na & 0 & \na & \na & 36 & 5.93 & 0.69\\ 
        SDL & 115 & 1 & 312 & 1 & 2 & \oldnew{0}{0} & \oldnew{0}{0} & 0 & 0 & \na & \na & 26 & 10.0 & 0.21\\ 
        NanoNNG & 268 & 1 & 16 & 5 & 5 & \oldnew{0}{0} & \oldnew{0}{0} & 0 & 2 & \na & \na & 142 & 6.66 & 1.25\\ 
        krb5 & 242 & 1 & 186 & 0 & 0 & \na & \na & \na & \na & \na & \na & \na & \na & \na\\ 
        opendmarc & 292 & 1 & 40 & 5 & 5 & \oldnew{0}{0} & \oldnew{0}{0} & 0 & 0 & \na & \na & 13 & 1.94 & 0.13\\ 
        Tcpreplay & 39.1 & 1 & 35 & 3 & 2 & \oldnew{3}{0} & \oldnew{1}{1} & 0 & 1 & \saneval{100} & \saneval{100} & 10.3 & 1.38 & 0.22\\
        yasm & 67.4 & 2 & 774 & 4 & 1 & \oldnew{0}{0} & \oldnew{0}{0} & 0 & 0 & \na & \na & 123 & 20.4 & 0.004\\ 
        cblosc2 & 98.0 & 4 & 114 & 0 & 1 & \na & \oldnew{0}{0} & \na & 0 & \na & \saneval{100} & 59 & 11.3 & 0.58\\
        libyang & 101 & 1 & 30 & 0 & 0 & \na & \na & \na & \na & \na & \na & \na & \na & \na\\
        openSSL & 465 & 6 & 27 & 15 & 10 & \oldnew{6}{3} & \oldnew{6}{3} & 6 & 0 & \saneval{100} & \saneval{89} & 394 & 25.3 & 0.87\\
        \midrule
        Overall & {\bf 3002.5} & {\bf 49}  & & 86 & 84 & \oldnew{33}{19} & \oldnew{28}{18} & 59 & 94 &  &   & 2298.5 &  \\
        Relative \% & & & & 35\% & 34\% & {\bf 52\%} & {\bf 55\%} &    &   & &  \\
        \bottomrule
    \end{tabularx}
\end{table*}

\begin{table*}[htbp]
    \footnotesize
    \centering
    \caption{Unbiased Dataset - statistics for experiments on 5 projects comprising 53 subjects with no known issues. The meaning of the columns are the same as those for \autoref{tab:CVE}, with the exception of \emph{Manifest Bugs} which only reports new bugs.} \label{tab:autogen}
    \begin{tabularx}{\textwidth}{lYY*{2}{Y}*{2}{Y}*{2}{Y}*{2}{Y}*{2}{Y}*{2}{Y}*{2}{Y}}
        \toprule
        \multirow{2}{*}{\emph{Project}} & \multirow{1}{*}{\emph{kLoC}} & \multirow{1}{*}{\emph{\#Fnc}} & \multirow{1}{*}{\emph{Avg.}} & \multicolumn{2}{c}{\emph{Syntactically Valid}} &\multicolumn{2}{c}{\emph{Manifest Bugs}} & \multicolumn{2}{c}{\emph{Latent Bugs}} & \multicolumn{2}{c}{\emph{Sanitised}} & \multicolumn{3}{c}{\emph{Average Time(s)}} \\
        \cmidrule(l){5-6}  \cmidrule(l){7-8} \cmidrule(l){9-10} \cmidrule(l){11-12} \cmidrule(l){13-15}
        &  & \multirow{1}{*}{}& \multirow{1}{*}{\emph{LoC}} & 3.5 & 4o & 3.5 & 4o & 3.5 & 4o & 3.5 & 4o & \emph{Total} & \emph{Blame} & \emph{San}\\
        \midrule
        yasm & 67.4 & 2 & 38 & 1 & 0  & 0 & \na & 0 & \na & \na & \na & 126 & 20.5 & 0.10\\ 
        NanoNNG & 268 & 3 & 7 & 10 & 11 & \manifest{10} & \manifest{11} & 0 & 0 & \saneval{100} & \saneval{100} & 156 & 6.69 & 1.47\\ 
        cblosc2 & 98.0 & 1 & 21 & 5 & 5 & \manifest{0} & \manifest{0} & 0 & 0 & \na & \na & 62.3 & 11.2 & 0.28\\
        gpac & 690 & 34 & 25 & 141 & 144 & \manifest{50} & \manifest{42} & 82 & 88 & \saneval{92} & \saneval{88} & 691 & 77.2 & 1.02\\ 
        openSSL & 465 & 13 & 15 & 49 & 58 & \manifest{24} & \manifest{22} & 17 & 19 & \saneval{75} & \saneval{73} & 394 & 28.1 & 0.77\\
        \midrule
        Overall & {\bf 1588.4} & {\bf 53} & &  211 & 218 &   \manifest{84} & \manifest{75} & 99 & 107 &   &  & 1429.3  &  \\
        Relative \% & & & & 80\% & 82\%  & {\bf 40\%} & {\bf 34\%} &  & &  &\\
        \bottomrule
    \end{tabularx}
    \vspace{-0.7em}
\end{table*}


\subsection{RQ2 - is LLM introducing unsafety issues?}
Asking \gpt\, to regenerate functions with historically known safety issues might bias the code generation towards unsafe code, even if the function has since been patched. To mitigate this potential bias, we repeat the previous experiment but now on 53 new subjects with no previously known safety issues, totalling 530 auto-generated functions. To confirm the bugs, we  use both \pulse\, and manually compare the manipulation of pointers in the auto-generated code against their manipulation in the original safe function.

\noindent\emph{\bf Results}. \autoref{tab:autogen} summarises our validation results. About 80\% and 82\%, respectively, of the auto-generated functions are syntactically valid. Among these, \gpt, introduced bugs that could manifest into issues in about 40\% and 34\% of the cases, respectively. The rate of syntactically valid code is higher than in the previous experiment, as well as the rate of the safe code produced. 
This could possibly be explained by: (1) the absence of unsafety bias in the training data, and (2) the smaller size of subjects in the unbiased dataset.
Although smaller rate of unsafe code, we consider it significantly high to require sanitisation intervention. 
 The distribution of manifesting bug types for this experiment is illustrated in \autoref{chart:auto-gen}. 

\begin{tcolorbox}[blanker,top=3mm,bottom=3mm,borderline horizontal={2pt}{0pt}{gray}]
\emph{Even for code with no prior safety issues, \gpt\, still introduces unsafe conditions, with about 28\% to 32\% of these conditions potentially becoming manifesting errors.}
\end{tcolorbox}

\subsection{RQ3 - is \tool\,effective in regaining safety?}
We ran \tool\, according to the pipeline in \autoref{fig:silc} for each syntactically valid auto-generated function. We assess whether our recipe for assigning blame for the manifesting bug and applying corresponding sanitization is sufficient to regain the project's safety after the integration of the auto-generated code.  

\noindent\emph{\bf Results}. The \emph{Sanitised} columns in \autoref{tab:CVE} and \autoref{tab:autogen} indicate that \tool\, regains safety for the majority of considered subjects. For few subjects where \tool\, did not regain safety, the fault lies within the bug detection phase. \tool\, relies on \pulse\, for detecting unsafe memory manipulation. \pulse\, is sound for bug detection, i.e. for each reported bug there is a proof to indicate its presence. However, \pulse\, has no guarantee to discover all bugs. The cases where \tool\, did not regain safety are those in which \pulse\, generated no bug report, yet our manual investigation indicates that a memory safety bug does exist. Performance wise, \tool\, spends the most amount analysing the code, which, as expected, is proportional to the project size. For \mcode{gpac}, the largest project in our datasets, once a function has been regenerated, it takes no more than 12 minutes to complete an end-to-end analysis, including the analysis of the entire project, identifying  blame and the sanitization of the auto-generated function. The analysis overhead introduced by adding support for blame-carrying proofs is of 2 minutes for the largest project, while the time for generating a sanitizer is kept below 2 seconds. 
\begin{tcolorbox}[blanker,top=3mm,bottom=3mm,borderline horizontal={2pt}{0pt}{gray}]
\emph{\tool\, requires a proof for the existence of the manifesting bugs. It sanitises 100\% of the bugs which do have a proof.}
\end{tcolorbox}

\subsection{Observations}
Manually examining the auto-generated functions, we noticed that \gptfour\, thoroughly guards pointer dereferences, even when not required. This could explain why the incidence of NPD is lower than that of ML for both datasets. However, reducing the incidence of ML during integration is more challenging without interventions like those added by \tool, because it would require \gpt\, to have in-depth understanding of program semantics and memory manipulation. Also, reducing the incidence of NPD which results from how the client uses the possible NULL pointers returned by the auto-generated code may also require such an external intervention. 

We also observed a correlation between safety  and the similarity between the code in the preamble and the generated code. If the prompt contains functions with similar functionality to the generated code, \gpt\, adopts the memory safety protocol from the preamble, thus leading to safer code.

It is unclear what memory safety practices one-shot \gpt\, follows. What our experiments confirm is that the rate of auto-generated code which could introduce memory unsafety when integrated into existing codebases is significant, and its use should be treated cautiously. 

\subsection{Threats to Validity}

One concern is the choice of subjects, whether they are more prone to producing unsafe code than other similar projects. We chose mature projects with an average of over 2k GitHub stars.
We also varied the subjects in these projects, including those with known issues (\autoref{tab:CVE}) and those with no previously known issues (\autoref{tab:autogen}). 
Another concern is 
whether a better prompt would reduce the rate of unsafe conditions produced by \gpt. We chose to follow the guidelines for prompt engineering offered by OpenAI \cite{openAI}, rather than delve deep into the research area of prompt engineering which has its own merits but it is orthogonal to our goals. 
Another threat to the validity is the randomness in the experiments due to the reliance on LLM to generate code. To mitigate this, each function has been generated 5 times, and while the reports on unsafe code may be noisy, the sanitisation success rate remains constant. 

Finally, \tool\, relies on \pulse for proving the existence of memory safety bugs. Therefore, our evaluation results are sensitive to the soundness of \pulse\,which is proved to be sound in theory \cite{Le2022}. To mitigate potential implementation issues, we also manually validated all the evaluation results.

\section{Related Work}
\label{sec:related_work}
\noindent \emph{\bf Auto-generated code.} 
Previous research highlights the need to focus on the non-functional properties of auto-generated code, highlighting issues that are insufficiently explored, such as safety, privacy, efficiency, and usability \cite{LLM4Code}.
Our work does a step in this direction, and to the best of our knowledge it is the first to be looking into the memory safety issues which arise as a result of \emph{integrating} auto-generated code into existing codebases.
A detailed study in a synthetic, micro scale coding environments found that about 40\% of code generated by GitHub's Copilot for languages such as C, Python, and Verilog contains vulnerabilities \cite{Hammond2022}.  
Our work goes beyond synthetic benchmarks, being the first to offer insights into the safety of auto-generated C code for \emph{real world projects}, a setup known to harbour notoriously disruptive bugs \cite{crowdstrike,heartbleed}.  
While existing research for safer auto-generated code primarily focuses on finetunned models to produce safer code \cite{He2023,olausson2023demystifying,Codexity2024,tihanyi2024newerasoftwaresecurity} our blame shifting framework is orthogonal being amenable to any sort of oracle for auto-generated code. 

\emph{\bf A blame logic.} Separation Logic based tools have shown promising results in repairing unsafe code \cite{TonderG18,He2023,song2024provenfix}. However, most of these repairing techniques do require user input when specifying the repair location, or default to the imprecise static analysis report. As shown in the motivating example, the state-of-the-art analysis report is not precise enough when dealing with complex call chains. While we do not target the code repair, a blame-carrying ISL could be explored in the future for better repair localization. 

Our work is also related to the body of work on higher-order contract enforcement with blame assignment for contract violation \cite{blamecontracts}. A Functional Perl offers an overview on this line of work \cite{blamefunctionalperl}. However, there are three main aspects which sets our work on a different trajectory from that on higher-order contracts. 
First, until recent work conducted simultaneously with ours \cite{popl2024}, these studies did not support reasoning about effects, being focused on the functional properties of target programs. Second, these blame contracts are mostly studied in the context of functional languages, while we target imperative C programs which require a different kind of reasoning due to the heavy manipulation of pointers. And third, these contracts are enforced dynamically, while ours is a static approach with no runtime interference.



\section{Perspectives}

We empirically demonstrate that the memory unsafety introduced by LLMs is significant, and provide examples showing that identifying the root causes of these safety issues can be challenging, when auto-generated code is integrated into existing codebases. 
%
To address this, we design \tool, a blame-shifting framework with sanitization capabilities grounded in what we call \emph{blame-carrying logic}. We show its effectiveness in assigning blame for safety violation and in automatically disabling unsafe paths resulted from integration. 
Our compositional treatment of blame-carrying proofs follows closely that of ISL, so that blame can be inferred from local properties established independently for all program components. 
\tool\ highlights the potential role of Incorrectness Logic in ensuring \emph{harmonious integration} of code from LLMs into software projects.
%
%
%
As the community works towards generating safer LLM code, this work emphasizes the trust boundaries that lie between manually written code and automatically generated code. The blame shifting framework allows us to accurately define and navigate  these trust boundaries.

%

\noindent \emph{\bf Data Availability}
The artifact accompanying this paper is available online \cite{silc}.
It contains the source code of \tool, the two datasets we collected to assess the safety of auto-generated code,
and scripts to reproduce our evaluation results.

\bibliographystyle{IEEEtran}
\bibliography{bibliography}

\begin{thebibliography}{10}
\providecommand{\url}[1]{#1}
\csname url@samestyle\endcsname
\providecommand{\newblock}{\relax}
\providecommand{\bibinfo}[2]{#2}
\providecommand{\BIBentrySTDinterwordspacing}{\spaceskip=0pt\relax}
\providecommand{\BIBentryALTinterwordstretchfactor}{4}
\providecommand{\BIBentryALTinterwordspacing}{\spaceskip=\fontdimen2\font plus
\BIBentryALTinterwordstretchfactor\fontdimen3\font minus \fontdimen4\font\relax}
\providecommand{\BIBforeignlanguage}[2]{{%
\expandafter\ifx\csname l@#1\endcsname\relax
\typeout{** WARNING: IEEEtran.bst: No hyphenation pattern has been}%
\typeout{** loaded for the language `#1'. Using the pattern for}%
\typeout{** the default language instead.}%
\else
\language=\csname l@#1\endcsname
\fi
#2}}
\providecommand{\BIBdecl}{\relax}
\BIBdecl

\bibitem{githubCopilot}
{GitHub}, ``{GitHub Copilot},'' \url{https://copilot.github.com/}, 2021.

\bibitem{Haonan2024}
\BIBentryALTinterwordspacing
H.~Li, Y.~Hao, Y.~Zhai, and Z.~Qian, ``Enhancing static analysis for practical bug detection: An llm-integrated approach,'' \emph{Proc. ACM Program. Lang.}, vol.~8, no. OOPSLA1, apr 2024. [Online]. Available: \url{https://doi.org/10.1145/3649828}
\BIBentrySTDinterwordspacing

\bibitem{Sun2024}
\BIBentryALTinterwordspacing
Y.~Sun, D.~Wu, Y.~Xue, H.~Liu, H.~Wang, Z.~Xu, X.~Xie, and Y.~Liu, ``Gptscan: Detecting logic vulnerabilities in smart contracts by combining gpt with program analysis,'' in \emph{Proceedings of the IEEE/ACM 46th International Conference on Software Engineering}, ser. ICSE '24.\hskip 1em plus 0.5em minus 0.4em\relax New York, NY, USA: Association for Computing Machinery, 2024. [Online]. Available: \url{https://doi.org/10.1145/3597503.3639117}
\BIBentrySTDinterwordspacing

\bibitem{meng2024large}
R.~Meng, M.~Mirchev, M.~B{\"o}hme, and A.~Roychoudhury, ``Large language model guided protocol fuzzing,'' in \emph{Proceedings of the 31st Annual Network and Distributed System Security Symposium (NDSS)}, 2024.

\bibitem{fan2023automated}
Z.~Fan, X.~Gao, M.~Mirchev, A.~Roychoudhury, and S.~H. Tan, ``Automated repair of programs from large language models,'' in \emph{2023 IEEE/ACM International Conference on Software Engineering (ICSE)}, 2023.

\bibitem{Chunqiu2023}
C.~S. Xia, Y.~Wei, and L.~Zhang, ``Automated program repair in the era of large pre-trained language models,'' in \emph{2023 IEEE/ACM 45th International Conference on Software Engineering (ICSE)}, 2023, pp. 1482--1494.

\bibitem{Joshi2023}
\BIBentryALTinterwordspacing
H.~Joshi, J.~C. Sanchez, S.~Gulwani, V.~Le, I.~Radi\v{c}ek, and G.~Verbruggen, ``Repair is nearly generation: multilingual program repair with llms,'' ser. AAAI'23/IAAI'23/EAAI'23.\hskip 1em plus 0.5em minus 0.4em\relax AAAI Press, 2023. [Online]. Available: \url{https://doi.org/10.1609/aaai.v37i4.25642}
\BIBentrySTDinterwordspacing

\bibitem{szafraniec2023code}
\BIBentryALTinterwordspacing
M.~Szafraniec, B.~Roziere, H.~J. Leather, P.~Labatut, F.~Charton, and G.~Synnaeve, ``Code translation with compiler representations,'' in \emph{The Eleventh International Conference on Learning Representations}, 2023. [Online]. Available: \url{https://openreview.net/forum?id=XomEU3eNeSQ}
\BIBentrySTDinterwordspacing

\bibitem{Baptiste2020}
B.~Roziere, M.-A. Lachaux, L.~Chanussot, and G.~Lample, ``Unsupervised translation of programming languages,'' in \emph{Proceedings of the 34th International Conference on Neural Information Processing Systems}, ser. NIPS '20.\hskip 1em plus 0.5em minus 0.4em\relax Red Hook, NY, USA: Curran Associates Inc., 2020.

\bibitem{eniser2024translating}
\BIBentryALTinterwordspacing
H.~F. Eniser, H.~Zhang, C.~David, M.~Wang, M.~Christakis, B.~Paulsen, J.~Dodds, and D.~Kroening, ``Towards translating real-world code with llms: A study of translating to rust,'' 2024. [Online]. Available: \url{https://arxiv.org/abs/2405.11514}
\BIBentrySTDinterwordspacing

\bibitem{puri2021codenet}
R.~Puri, D.~Kung, A.~Gholami, Q.~Yu, N.~Mellempudi, E.~Hendrickson, Q.~Huang, A.~Krishnamurthy, D.~Sitaram, D.~Das, D.~Mudigere, D.~Dutta, G.~Taylor, P.~van~der Smagt, H.~Yuen, S.~Pankanti, S.~Ohmer, S.~Park, T.~Kanter, W.~Tang, W.~Smach, and Y.~Suzuki, ``Codenet: A large-scale ai for code dataset for learning a diversity of coding tasks,'' \emph{arXiv preprint arXiv:2105.12655}, 2021.

\bibitem{CopilotImpact2024}
\BIBentryALTinterwordspacing
A.~Ziegler, E.~Kalliamvakou, X.~A. Li, A.~Rice, D.~Rifkin, S.~Simister, G.~Sittampalam, and E.~Aftandilian, ``{Measuring GitHub Copilot's Impact on Productivity},'' \emph{Commun. ACM}, vol.~67, no.~3, p. 54–63, feb 2024. [Online]. Available: \url{https://doi.org/10.1145/3633453}
\BIBentrySTDinterwordspacing

\bibitem{AIProgrammingAssistants2024}
\BIBentryALTinterwordspacing
J.~T. Liang, C.~Yang, and B.~A. Myers, ``A large-scale survey on the usability of ai programming assistants: Successes and challenges,'' in \emph{Proceedings of the IEEE/ACM 46th International Conference on Software Engineering}, ser. ICSE '24.\hskip 1em plus 0.5em minus 0.4em\relax New York, NY, USA: Association for Computing Machinery, 2024. [Online]. Available: \url{https://doi.org/10.1145/3597503.3608128}
\BIBentrySTDinterwordspacing

\bibitem{ICSEcopilot2022}
\BIBentryALTinterwordspacing
S.~Imai, ``{Is GitHub copilot a substitute for human pair-programming? an empirical study},'' in \emph{Proceedings of the ACM/IEEE 44th International Conference on Software Engineering: Companion Proceedings}, ser. ICSE '22.\hskip 1em plus 0.5em minus 0.4em\relax New York, NY, USA: Association for Computing Machinery, 2022, p. 319–321. [Online]. Available: \url{https://doi.org/10.1145/3510454.3522684}
\BIBentrySTDinterwordspacing

\bibitem{GILT2024}
\BIBentryALTinterwordspacing
D.~Nam, A.~Macvean, V.~Hellendoorn, B.~Vasilescu, and B.~Myers, ``Using an llm to help with code understanding,'' in \emph{Proceedings of the IEEE/ACM 46th International Conference on Software Engineering}, ser. ICSE '24.\hskip 1em plus 0.5em minus 0.4em\relax New York, NY, USA: Association for Computing Machinery, 2024. [Online]. Available: \url{https://doi.org/10.1145/3597503.3639187}
\BIBentrySTDinterwordspacing

\bibitem{expectations2022}
\BIBentryALTinterwordspacing
P.~Vaithilingam, T.~Zhang, and E.~L. Glassman, ``{Expectation vs. Experience: Evaluating the Usability of Code Generation Tools Powered by Large Language Models},'' in \emph{Extended Abstracts of the 2022 CHI Conference on Human Factors in Computing Systems}, ser. CHI EA '22.\hskip 1em plus 0.5em minus 0.4em\relax New York, NY, USA: Association for Computing Machinery, 2022. [Online]. Available: \url{https://doi.org/10.1145/3491101.3519665}
\BIBentrySTDinterwordspacing

\bibitem{Jiawei2024}
J.~Liu, C.~S. Xia, Y.~Wang, and L.~Zhang, ``Is your code generated by chatgpt really correct? rigorous evaluation of large language models for code generation,'' in \emph{Proceedings of the 37th International Conference on Neural Information Processing Systems}, ser. NIPS '23.\hskip 1em plus 0.5em minus 0.4em\relax Red Hook, NY, USA: Curran Associates Inc., 2024.

\bibitem{misu2024towards}
M.~R.~H. Misu, C.~V. Lopes, I.~Ma, and J.~Noble, ``Towards ai-assisted synthesis of verified dafny methods,'' \emph{arXiv preprint arXiv:2402.00247}, 2024.

\bibitem{CyberSecEval2}
\BIBentryALTinterwordspacing
M.~Bhatt, S.~Chennabasappa, Y.~Li, C.~Nikolaidis, D.~Song, S.~Wan, F.~Ahmad, C.~Aschermann, Y.~Chen, D.~Kapil, D.~Molnar, S.~Whitman, and J.~Saxe, ``{CyberSecEval 2: A Wide-Ranging Cybersecurity Evaluation Suite for Large Language Models},'' 2024. [Online]. Available: \url{{https://arxiv.org/abs/2404.13161}}
\BIBentrySTDinterwordspacing

\bibitem{surveysecurity2023}
\BIBentryALTinterwordspacing
N.~Perry, M.~Srivastava, D.~Kumar, and D.~Boneh, ``Do users write more insecure code with ai assistants?'' in \emph{Proceedings of the 2023 ACM SIGSAC Conference on Computer and Communications Security}, ser. CCS '23.\hskip 1em plus 0.5em minus 0.4em\relax New York, NY, USA: Association for Computing Machinery, 2023, p. 2785–2799. [Online]. Available: \url{https://doi.org/10.1145/3576915.3623157}
\BIBentrySTDinterwordspacing

\bibitem{Raad2020}
A.~Raad, J.~Berdine, H.-H. Dang, D.~Dreyer, P.~O'Hearn, and J.~Villard, ``Local reasoning about the presence of bugs: Incorrectness separation logic,'' in \emph{Computer Aided Verification}, S.~K. Lahiri and C.~Wang, Eds.\hskip 1em plus 0.5em minus 0.4em\relax Cham: Springer International Publishing, 2020, pp. 225--252.

\bibitem{Le2022}
\BIBentryALTinterwordspacing
Q.~L. Le, A.~Raad, J.~Villard, J.~Berdine, D.~Dreyer, and P.~W. O'Hearn, ``Finding real bugs in big programs with incorrectness logic,'' \emph{Proc. ACM Program. Lang.}, vol.~6, no. OOPSLA1, apr 2022. [Online]. Available: \url{https://doi.org/10.1145/3527325}
\BIBentrySTDinterwordspacing

\bibitem{openNDS}
``{openNDS},'' \url{https://github.com/openNDS/openNDS}, 2024.

\bibitem{memleakopenNDS}
``{CVE-2023-41102},'' \url{https://cve.mitre.org/cgi-bin/cvename.cgi?name=CVE-2023-41102}, 2023.

\bibitem{MemLock}
\BIBentryALTinterwordspacing
C.~Wen, H.~Wang, Y.~Li, S.~Qin, Y.~Liu, Z.~Xu, H.~Chen, X.~Xie, G.~Pu, and T.~Liu, ``{MemLock: memory usage guided fuzzing},'' ser. ICSE '20.\hskip 1em plus 0.5em minus 0.4em\relax New York, NY, USA: Association for Computing Machinery, 2020, p. 765–777. [Online]. Available: \url{https://doi.org/10.1145/3377811.3380396}
\BIBentrySTDinterwordspacing

\bibitem{gpac}
``{GPAC},'' \url{https://github.com/gpac/gpac}, 2024.

\bibitem{npdGPAC}
``{CVE-2024-6062},'' \url{https://cve.mitre.org/cgi-bin/cvename.cgi?name=CVE-2024-6062}, 2024.

\bibitem{HumanEval}
\BIBentryALTinterwordspacing
M.~Chen, J.~Tworek \emph{et~al.}, ``{Evaluating Large Language Models Trained on Code},'' \emph{CoRR}, vol. abs/2107.03374, 2021. [Online]. Available: \url{https://arxiv.org/abs/2107.03374}
\BIBentrySTDinterwordspacing

\bibitem{MBPP}
\BIBentryALTinterwordspacing
J.~Austin, A.~Odena \emph{et~al.}, ``Program synthesis with large language models,'' \emph{CoRR}, vol. abs/2108.07732, 2021. [Online]. Available: \url{https://arxiv.org/abs/2108.07732}
\BIBentrySTDinterwordspacing

\bibitem{EvoCodeBench}
\BIBentryALTinterwordspacing
J.~Li, G.~Li, X.~Zhang, Y.~Dong, and Z.~Jin, ``{EvoCodeBench: An Evolving Code Generation Benchmark Aligned with Real-World Code Repositories},'' 2024. [Online]. Available: \url{https://arxiv.org/abs/2404.00599}
\BIBentrySTDinterwordspacing

\bibitem{cve}
``{CVE},'' \url{https://cve.mitre.org/}.

\bibitem{openAI}
{OpenAI}, ``{Best practices for prompt engineering with the OpenAI API},'' \url{https://help.openai.com/en/articles/6654000-best-practices-for-prompt-engineering-with-the-openai-api}, 2024.

\bibitem{LLM4Code}
\BIBentryALTinterwordspacing
Z.~Yang, Z.~Sun, T.~Z. Yue, P.~Devanbu, and D.~Lo, ``{Robustness, Security, Privacy, Explainability, Efficiency, and Usability of Large Language Models for Code},'' 2024. [Online]. Available: \url{https://arxiv.org/abs/2403.07506}
\BIBentrySTDinterwordspacing

\bibitem{Hammond2022}
H.~Pearce, B.~Ahmad, B.~Tan, B.~Dolan-Gavitt, and R.~Karri, ``{Asleep at the Keyboard? Assessing the Security of GitHub Copilot’s Code Contributions},'' in \emph{2022 IEEE Symposium on Security and Privacy (SP)}, 2022, pp. 754--768.

\bibitem{crowdstrike}
{CrowdStrike}, ``{Preliminary Post Incident Review (PIR): Content Configuration Update Impacting the Falcon Sensor and the Windows Operating System (BSOD)},'' \url{https://www.crowdstrike.com/falcon-content-update-remediation-and-guidance-hub/}, 2024.

\bibitem{heartbleed}
``{The Heartbleed Bug},'' \url{https://heartbleed.com}, 2014.

\bibitem{He2023}
\BIBentryALTinterwordspacing
J.~He and M.~Vechev, ``Large language models for code: Security hardening and adversarial testing,'' in \emph{Proceedings of the 2023 ACM SIGSAC Conference on Computer and Communications Security}, ser. CCS '23.\hskip 1em plus 0.5em minus 0.4em\relax New York, NY, USA: Association for Computing Machinery, 2023, p. 1865–1879. [Online]. Available: \url{https://doi.org/10.1145/3576915.3623175}
\BIBentrySTDinterwordspacing

\bibitem{olausson2023demystifying}
T.~X. Olausson, J.~P. Inala, C.~Wang, J.~Gao, and A.~Solar-Lezama, ``Demystifying gpt self-repair for code generation,'' \emph{arXiv preprint arXiv:2306.09896}, 2023.

\bibitem{Codexity2024}
\BIBentryALTinterwordspacing
S.~Y. Kim, Z.~Fan, Y.~Noller, and A.~Roychoudhury, ``Codexity: Secure ai-assisted code generation,'' 2024. [Online]. Available: \url{https://arxiv.org/abs/2405.03927}
\BIBentrySTDinterwordspacing

\bibitem{tihanyi2024newerasoftwaresecurity}
\BIBentryALTinterwordspacing
N.~Tihanyi, R.~Jain \emph{et~al.}, ``{A New Era in Software Security: Towards Self-Healing Software via Large Language Models and Formal Verification},'' 2024. [Online]. Available: \url{https://arxiv.org/abs/2305.14752}
\BIBentrySTDinterwordspacing

\bibitem{TonderG18}
R.~{van Tonder} and C.~{Le Goues}, ``Static automated program repair for heap properties,'' in \emph{ICSE}.\hskip 1em plus 0.5em minus 0.4em\relax {ACM}, 2018, pp. 151--162.

\bibitem{song2024provenfix}
Y.~SONG, X.~GAO, W.~LI, W.-N. CHIN, and A.~ROYCHOUDHURY, ``Provenfix: Temporal property guided program repair,'' to appear 2024.

\bibitem{blamecontracts}
\BIBentryALTinterwordspacing
R.~B. Findler and M.~Felleisen, ``Contracts for higher-order functions,'' \emph{SIGPLAN Not.}, vol.~37, no.~9, p. 48–59, sep 2002. [Online]. Available: \url{https://doi.org/10.1145/583852.581484}
\BIBentrySTDinterwordspacing

\bibitem{blamefunctionalperl}
\BIBentryALTinterwordspacing
C.~Dimoulas, M.~S. New, R.~B. Findler, and M.~Felleisen, ``Oh lord, please don't let contracts be misunderstood (functional pearl),'' \emph{SIGPLAN Not.}, vol.~51, no.~9, p. 117–131, sep 2016. [Online]. Available: \url{https://doi.org/10.1145/3022670.2951930}
\BIBentrySTDinterwordspacing

\bibitem{popl2024}
\BIBentryALTinterwordspacing
C.~Moy, C.~Dimoulas, and M.~Felleisen, ``Effectful software contracts,'' vol.~8, no. POPL, jan 2024. [Online]. Available: \url{https://doi.org/10.1145/3632930}
\BIBentrySTDinterwordspacing

\bibitem{silc}
``{SILC},'' \url{https://zenodo.org/uploads/13196914}, 2024.

\end{thebibliography}

\end{document}